\documentclass{article}
\usepackage[utf8]{inputenc}
\usepackage{graphicx} 
\usepackage{amsmath}

\usepackage{amssymb}
\usepackage{array}

\newtheorem{theorem}{Theorem}

\newtheorem{definition}[theorem]{Definition}
\newtheorem{proposition}[theorem]{Proposition}

\newtheorem{remark}[theorem]{Remark}
\newtheorem{example}[theorem]{Example}
\newtheorem{question}[theorem]{Question}
\newtheorem{opinion}{Take home message}
\newenvironment{proof}{\noindent\bf{Proof.}\rm}{\hfill$\blacksquare$\bigskip}

\newcommand{\items}{\mathcal{M}} 
\newcommand{\agents}{\mathcal{N}} 

\usepackage[]{color-edits}
\addauthor{UF}{red}
\newcommand{\ufc}[1]{{\UFcomment{#1}}}

\addauthor{GB}{blue}

\title{The inversion paradox, and classification of fairness notions}
\author{Uriel Feige\thanks{Weizmann Institute, Israel. {\tt uriel.feige@weizmann.ac.il}}}

\begin{document}

\maketitle

\begin{abstract}
Several different fairness notions have been introduced in the context of fair allocation of goods. In this manuscript, we compare between some fairness notions that are used in settings in which agents have arbitrary (perhaps unequal) entitlements to the goods. This includes the proportional share, the anyprice share, the weighted maximin share, weighted envy freeness, maximum weight Nash social welfare and competitive equilibrium. We perform this comparison in two settings, that of a divisible homogeneous good and arbitrary valuations, and that of indivisible goods and additive valuations.

Different fairness notions are not always compatible with each other, and might dictate selecting different allocations.
The purpose of our work is to clarify various properties of fairness notions, so as to allow, when needed, to make an educated choice among them. Also, such a study may motivate introducing new fairness notions, or modifications to existing fairness notions.

Among other properties, we introduce definitions for monotonicity that postulate that having higher entitlement should be better to the agent than having lower entitlement. Some monotonicity notions, such as population monotonicity and weight monotonicity, appeared in previous work, but we prefer to consider other monotonicity properties that we refer to as global monotonicity and individual monotonicity. We find that some of the fairness notions (but not all) violate our monotonicity properties in a strong sense, that we refer to as the inversion paradox. Under this paradox, a fairness notion enforces that the value received by an agent decreases when the entitlement of the agent increases.  
\end{abstract}

\section{Introduction}

In this manuscript we consider allocation of goods to agents. In the setting that we consider there is either one item or several items to be allocated. The set of items is denoted by $\items$. There is a set $\agents$ of $n \ge 2$ agents that may receive items. Items are considered to be {\em goods}, in the sense that agents wish to receive them (as opposed to {\em bads} or {\em chores}, that agents prefer not to receive). Items may be either divisible (agents may receive fractions of a divisible item) or indivisible (an agent may either receive the item as a whole, or no part of it). An allocation $A = (A_1, \ldots, A_n)$ is a partition $\items$ among the $n$ agents -- each indivisible item is allocated to one agent, and for each divisible item, the fractions allocated to agents sum up to~1. Each part $A_i$, received by agent $i$, will be referred to as a bundle of items (and may include also fractions of divisible items). In our setting, agents do not pay in order to receive their bundle. Rather, prior to the allocation, agents have initial entitlements to the set of items, and following the allocation, agents give up their entitlements to those bundles that were received by others. Informally, an allocation is {\em fair} if no agent has a ``justifiable" complaint that she received a bundle of too small value.

Let us present an example to illustrate our setting. The example is that of dividing an inheritance. Suppose that the legal will of a person who died states that a $\frac{2}{5}$ fraction of his estate (everything that the person owned when he died) should go to his widowed wife, and each of his three children 
should get a $\frac{1}{5}$ fraction of his estate. In this case, $\items$ is the estate of the deceased person, the set $\agents$ is composed of the wife and the three children, and the entitlements are $(\frac{2}{5}, \frac{1}{5}, \frac{1}{5}, \frac{1}{5})$ (summing up to~1). Now let us delve deeper into this example, by examining the possible content of the estate.

\begin{itemize}
    \item The estate consists of only money, say, $M$ dollars. In this case we may think of $\items$ as composed of a single divisible item (the set of $M$ dollars). Moreover, it is a {\em homogeneous} good (each dollar is as good as any other dollar). In terms of the value of the good to the agents, it makes sense to enforce an external valuation function on all agents, which is simply linear in the number of dollars that they receive. In such a setting, all fairness notions that we consider in this paper suggest the {\em proportional allocation}: give the wife $\frac{2M}{5}$ dollars (if needed for this purpose, some dollars can be broken into cents), and give each of the children $\frac{M}{5}$ dollars. We refer to this setting as a setting with a homogeneous divisible good, with an external valuation function, and moreover, this external valuation function is linear.

    \item The estate consists of $M$ dollars, but according to the laws of the country, individuals receiving an inheritance can be taxed. The amount of tax to be paid is a certain percentage of the inheritance received, where this percentage depends on the amount inherited by the person, and perhaps also on additional factors that are specific to the taxed person (e.g., her income from other sources during the same calendar year). Now the value that an agent attributes to dollars allocated to her might be smaller than their face value (due to the need to pay taxes), and moreover, need not be linear (tax rates may differ depending on the amount received). Different agents might have different valuation functions (if their tax status is different). We are still in a setting with a homogeneous divisible good, but now there are individual valuation functions, and moreover, these valuation functions need not be linear. Should the proportional allocation be used also in this setting? If not, what other allocations are fair? As we shall see, different fairness notions introduced in previous literature suggest different answers to these questions. (Before reading on, readers are encouraged to come up with their own answers, as this may help them gauge how plausible they find the existing fairness notions.)

    \item The deceased person was a gifted painter with no interest in personal wealth, and the estate consists only of some of his paintings. Each painting is an indivisible item. Much of the value of the paintings is sentimental and subjective. Now we are in a setting in which $\items$ is a collection of indivisible items. Each agent has an individual valuation function (that depends on how much she likes each painting, and more generally, each subset of paintings). In this setting it seems more difficult to articulate what properties one would like fair allocations to have. In fact, as we shall see, most fairness notions considered in this paper are such that for some allocation instances, no allocation is considered to be fair (according to these notions).
   
\end{itemize}

Other examples of allocating a homogeneous divisible good include allocation of storage space (e.g., physical storage in a shared warehouse, or digital storage on a shared disk space), allocation of time (e.g., for medical tests on an MRI machine, or allocating CPU time to jobs), allocation of transmission bandwidth, sharing water resources, or a more benign example -- dividing a chocolate cake among members of the family. Other examples of allocating indivisible items include allocation of places in university courses (students at different stages of their studies might have different entitlements), allocation of basketball players to NBA teams in the NBA draft (there the entitlement of a team is negatively correlated with its performance in the previous season), and allocation of housing units to eligible residents. 

\subsection{Several fairness notions}

There is extensive literature on fair allocation, and many fairness notions have been defined. An allocation that satisfies a given fairness notion will be referred to as {\em acceptable} under the fairness notion. It is not practical to try to review all fairness notions that appear in the literature. Instead, we shall consider six of these fairness notions, which we view as a diverse sample of those fairness notions that apply to settings with possibly unequal entitlements. The fairness notions will be defined and discussed in Section~\ref{sec:notions}, and here we just list them for the benefit of those readers who are already familar with them. (The references given to each of the fairness notion refer to papers in which the fairness notions were used in settings with arbitrary entitlements.)

\begin{itemize}
    \item Share based fairness notions. An allocation is acceptable if it gives every agent at least her share value. We shall consider three different ways of defining share values.
    \begin{itemize}
        \item The {\em proportional share} Prop~\cite{Barbanel1996}.
        \item The {\em anyprice share} APS~\cite{BEF21APS}. 
        \item The {\em weighted maximin share} WMMS~\cite{FGHLPSSY19}.
    \end{itemize}
    \item Comparison based fairness notions. An allocation is acceptable if each agent finds that no other agent was treated more favorably that she was. 
    \begin{itemize}
        \item {\em Weighted envy freeness} WEF~\cite{RP98}.
    \end{itemize}
    \item Equilibrium based fairness notions. An allocation is acceptable if it is a best response of the agents to some hypothetical item prices. 
    \begin{itemize}
        \item {\em Competitive equilibrium} CE~\cite{BNT21}.
    \end{itemize}
    \item Optimization based fairness notions. An allocation is acceptable if it optimizes a given fairness objective.
    \begin{itemize}
        \item {\em Maximum weighted Nash social welfare} MWNSW~\cite{CISZ21}.
    \end{itemize}
\end{itemize}

As we shall see, some of the fairness notions are inconsistent with each other, in the sense that for some allocation instances the sets of acceptable allocations for different fairness notions are disjoint from each other. There may be several explanations to these inconsistencies, and we list here three plausible ones.

\begin{itemize}
    \item Possibly, people use the same word ``fair", but have a different interpretation of what this word means. The different fairness notions might reflect different social norms. 
    \item The mathematical formulation of allocation problems does not capture the real complexity of the situation.  For example, there might be a distinction between situations in which the items to be allocated are medical services that the agents need, or a cake that the agents wish to eat for pleasure. Likewise, there might be distinctions based on the attitudes that agents have towards each other (do they like each other or not? are they indifferent to each other?). The different fairness notions are meant to be applied in different situations.
    \item The research community is still in a phase of exploring various fairness definitions, the implications of these definitions are not yet fully understood, and there are misconceptions floating around. It could be that in the long run, some of the fairness notions will be more widely adopted, others will be used only rarely, and fewer inconsistencies will remain.  
\end{itemize}

As there are inconsistencies among fairness notions, the current paper attempts to clarify the implications of the fairness notions presented above, so as to allow for more educated choices among them (both in research, and in practice).

\subsection{One possible take home message}

Our paper presents many definitions, propositions and examples concerning properties that fairness notions have or fail to have. The author doubts that it is wise to attempt to compress the landscape painted by this paper to one short take home message. Nevertheless, we present here one possible such message, for the benefit of those readers who prefer to see one. 

Consider the following two statements.

\begin{enumerate}
    \item When allocating a divisible homogeneous good (such as money) among agents, the proportional allocation (that gives each agent a fraction of the good that is equal to her entitlement) is fair.
    \item More entitlement is better. Specifically, if an agent $j$ transfers some of her entitlement to an agent $i$, a fairness notion should not dictate that the receiving agent $i$ loses value. It should not be the case that the maximum possible value that agent $i$ can get after the transfer is strictly smaller than the minimum possible value that she might get without this transfer. 
\end{enumerate}

Among the six fairness notions considered in this paper, only APS and CE satisfy statement~1, and only APS and Prop satisfy statement~2. Consequently if one postulates that both statements should hold, then among the six fairness notions considered in this paper, only the APS is not disqualified from serving as a fairness notion.

We proceed to provide more details, offering a much more nuanced landscape than the one presented above.

\subsection{Entitlements and valuation functions}

For all fairness notions considered in this paper, the question of whether a given allocation is acceptable depends only on the entitlements of the agents and on their valuation functions. (In contrast, here are some aspects that our fairness notions do not depend on. One is the desires of the items. For example, when the agents are medical doctors and items are positions in hospitals, it may be that hospitals also have a ranking among the doctors. Another is attributes of the agents. For example, in allocating research grants to researchers, the researchers may be grouped based on attributes that they share, such as their gender, and a diversity fairness notion may require that at least a certain percentage of the grants are given to members of each group.) 

The entitlement of each agent $i$ is simply a fraction $0 < b_i < 1$, with $\sum_{i \in \agents} b_i = 1$. If $b_i = \frac{1}{n}$ for every agent $i$, then we refer to the allocation instance as one with equal entitlements. Other cases are referred to in previous work as instances with unequal entitlements, or as instances with arbitrary entitlements (allowing equal entitlement as a special case), or as weighted instances (as in the fairness notions of weighted maximin share and weighted envyfreeness). 

The valuation function $v_i$ of each agent $i$ specifies how much she values each bundle of items. For indivisible items, the bundle specifies which items are present.   For divisible items, the bundle specifies what fraction of the divisible item is present. Specifically, for a homogeneous divisible good, $v_i$ is function of one argument $x$ that denotes the fraction of the good received by the agent (that is $0 \le x \le 1$). We use the following conventions.

\begin{itemize}
    \item Normalization: the value of the empty bundle is~0.
    \item Monotonicity: if $S \subset T$ then $v_i(S) \le v_i(T)$. Here, for a divisible item $e$, the $\subset$ notation means that the fraction of $e$ belonging to $S$ is not larger than the fraction of $e$ belonging to $T$. The monotonicity property reflects our assumption that the items are {\em goods}.
    \item Positivity: there is a bundle of positive value, $v_i(\items) > 0$.
    \item Homogeneity: the contribution of a fraction of a divisible item to the valuation depends only on the size of the fraction. To model a non-homogeneous divisible item (e.g., a chocolate cake with a cherry on top), we assume that it is decomposed into a collection of homogeneous divisible items (one containing only the chocolate cake, the other only the cherry). 
    \item Scale invariance: the valuation function only specifies the relative values of bundles, not an absolute value. This assumption dictates that we only consider fairness notions that are themselves scale invariant, meaning that the set of acceptable allocations does not change if a valuation function is scaled by a multiplicative positive constant. Indeed, the six fairness notions considered in this paper are scale invariant.
\end{itemize}


In general, each agent has her own individual valuation function. A special case is when all valuations are identical. This kind of coincidence might not be rare in practice, because it comes up when some {\em external valuation function} is accepted as de facto representing the ``true" values of the items. In the inheritance example given above, this may naturally happen when the estate consists only of money. It may also happen when the estate is composed of indivisible items, if each item has a known ``market value", and agents choose to adopt the market values of the items as their own valuations.

Valuation functions may belong to certain natural classes. For a divisible homogeneous good, natural classes include linear functions (e.g., $v(x) = x$), concave functions (e.g., $v(x) = \sqrt{x}$), and convex functions (e.g., $v(x) = x^2$). For indivisible goods (and also for a non-homogeneous divisible good, which we model as a collection of homogeneous divisible goods), a class that received much attention is that of additive valuations ($v(S) = \sum_{e\in S} v(e)$). Some of the fairness notions that we consider (such as Prop) were initially defined in settings with additive valuations, with no claim that they should be applicable also to settings with more general valuation functions.

\subsection{Feasibility and consistent selection rules}

The question of whether a fairness notion is always applicable is captured by the notion of {\em feasibility}.

\begin{definition}
    \label{def:feasible}
    A fairness notion is {\em feasible} for a class $C$ of valuations if  every instance in which all agent have valuations from class $C$ has an allocation that is acceptable under the fairness notion.
\end{definition}

It turns out that except for MWNSW (and also for MWNSW there is a caveat that will be explained in Section~\ref{sec:feasibility}), the other five fairness notions considered in this paper are not feasible for some classes of valuation functions.  For these fairness notions, in the case of indivisible goods, it is well known that non-feasibility holds even for additive valuations. For the case of a homogeneous divisible good, we observe that the situation is mixed (some fairness notions are always feasible and some are not). This last observation might come as a  mild surprise to those who assume that the only source of infeasibility for fair allocation is indivisibility of items.

Faced with the infeasibility issue, there is extensive work on approximate versions of the fairness notions, so as to handle cases in which they are not feasible. Though the question of what form of approximation would be a proper one is important, it is not the focus of our paper. Instead we take the following view concerning how fairness notions guide the selection of allocations. We assume that the choice of allocation is made by a {\em selection rule}. The selection rule is either deterministic or randomized. If it is deterministic it outputs a single allocation, and if it is randomized it outputs a distribution over allocations (or a single allocation sampled from this distribution). In this paper, we make no assumption regarding what the actual selection rule is, except for assuming that it is consistent with the underlying fairness notion, in the following sense.

\begin{itemize}
    \item If there are allocations that are acceptable under the fairness notion, an acceptable allocation must be selected. If the allocation rule is randomized, its distribution is supported only on acceptable allocations (though not necessarily on all acceptable allocations). 
    \item If no acceptable allocation exists, then the selection rule may output any allocation. We make no assumption regarding in what sense the chosen allocation approximates the fairness notion under consideration.
    
\end{itemize}

The focus of this paper is to study the properties of the acceptable allocations in those instances in which the fairness notion is feasible. This study can assist in answering the following question:

{\em For a given fairness notion, what are the implications of using a selection rule that is consistent with the fairness notion? }

If the implications are undesirable, this may indicate that the respective fairness notion should not be used.



\subsection{Monotonicity}

As noted above, there are several different fairness notions, and they are based on various different principles (share based, comparison based, and others). Is there some unifying principle that one would expect all fairness notions to adhere to? To the author it seems that the following should be such a principle:  for allocation of goods, having higher entitlement is at least as good (for the agent) as having lower entitlement. In this section we attempt to present rigorous definitions that capture this principle. (Some previous related definitions are discussed in Section~\ref{sec:related}.)

We assume a setting in which the set of items, the set of agents and their valuation functions are already fixed. The only thing that may vary is the entitlements of the agents (which are nonnegative and need to sum up to~1). For a given fairness notion, we consider the sets of acceptable allocations under various assignments of entitlements, and try to articulate what constraints do the entitlement values impose on the sets of acceptable allocations. 



We shall keep in mind the fact that fairness notions are not always feasible, and moreover, even when they are feasible, they do not necessarily dictate a unique allocation, but rather offer a set of acceptable allocations. Consequently, given a fairness notion, we associate with each allocation instance the set of allocations that are acceptable under this fairness notion. Our monotonicity properties are formulated in a way that does not assume that this set contains exactly a single allocation: they allow the set to have multiple allocations, and also allow it to be empty.

In general, we do not assume anything about which allocation is selected among those that are acceptable. However, for some fairness notions, specifically, those that are share based, it is natural to select an allocation from the {\em Pareto front} of the set of acceptable allocations.

\begin{definition}
\label{def:pareto}
    Given valuation functions for the agents, an allocation $A = (A_1,\ldots, A_n)$ {\em Pareto dominates} an allocation $A' = (A'_1,\ldots, A'_n)$ if for every agent $i$ it holds that $v_i(A_i) \ge v_i(A'_i)$, and at least for one agent, this inequality is strict. An allocation is Pareto optimal if no other allocation Pareto dominates it. Given a set $S$ of allocations, the {\em Pareto front} of $S$ is the set of those allocations in $S$ that are not dominated by any other allocation in $S$. 
\end{definition}

In this paper, our results concerning monotonicity (or absence of monotonicity) hold regardless of whether or not we require selection rules to select an allocation from the Pareto front of the acceptable allocations. 

For monotonicity we offer two sets of definitions. One concerns what we refer to as {\em internal monotonicity}, 
whereas the other concerns what we refer to as {\em external monotonicity}.
 
\begin{definition}
\label{def:internalMonotonicity}
    A fairness notion satisfies {\em internal monotonicity} if its acceptable allocations satisfy the following properties. For every allocation instance that contains two agents $i$ and $j$ with the same valuation (call it $v$), for every vector of entitlements $b = (b_1, \ldots, b_n)$ for which an acceptable allocation exists, the following holds. If $b_i \ge b_j$, then in every acceptable allocation $A = (A_1, \ldots, A_n)$ either $v(A_i) \ge v(A_j)$, or the allocation $A'$ that differs from $A$ by switching between $A_i$ and $A_j$ is also acceptable.
\end{definition}

Internal monotonicity is a property that is easy to satisfy, and all the fairness notions considered in this paper satisfy it. It is applicable only if several agents have the same valuation function, and offers no constraints otherwise. Hence the main qualitative property of monotonicy that we shall consider in this paper is that of {\em external monotonicity}, and we shall refer to it as {\em monotonicty} (without need of prefixing this term by {\em external}). 

The interpretation for monotonicity that we take is that if one or more agents are willing to transfer some of their entitlements to agent $i$, then it is in the best interest of $i$ to agree. Hence monotonicity will compare between the sets of acceptable allocations in two instances that differ only in their vectors of entitlements, and moreover, this difference is such that exactly one agent gains in entitlement (and as entitlements sum up to~1, one or more agents lose in entitlement). 

\begin{definition}
\label{def:improves}
    Vector $b$ of entitlements {\em $i$-improves} vector $b'$, denoted by $b >_i b'$, if $b_i > b'_i$ and (importantly) $b_j \le b'_j$ for all $j \not= i$.
\end{definition}

Though not required by Definition~\ref{def:improves}, it will be the case that in all explicit examples that we give with $b >_i b'$, there will be a single agent $j$ for which $b_j < b'_j$ (the definition allows for any number of such agents).

In the case that one or more of the agents have positive entitlement in $b'$ and~0 entitlement in $b$, then the allocation instance for $b$ has fewer agents than the one for $b'$. (Interestingly, such cases do occur in practice. For example, it is quite common (in Israel) that when one of the parents dies, children transfer their part of the entitlement to the inheritance to the surviving parent.) Though our definitions are intended to allow for such an option, we do not make use of it in examples presented in this paper.

Informally, monotonicity is intended to imply that if $b >_i b'$, then under $b$ agent $i$ should receive a bundle with at least as high value as under $b'$. One way of satisfying this intention is through the following notion of global monotonicity. 

\begin{definition}
\label{def:global}
    A fairness notion is {\em globally monotone} if there is a deterministic selection rule that is consistent with the fairness notion and enforces the following. For every two allocation instances that differ only in their vector of entitlements, if the vectors of entitlements satisfy $b >_i b'$ for some agent $i$, then the value (under $v_i$) of the bundle received by agent $i$ under entitlement $b$ is at least as large as that under entitlement $b'$.
\end{definition}

Definition~\ref{def:global} requires a selection rule that is deterministic.
In input instances that have symmetries (e.g., two agents with identical valuation functions), one is naturally led to use randomized allocation rules. For the purpose of removing randomness in such settings (so as to apply Definition~\ref{def:global}), and only for this purpose, we allow the selection rule to use names of the agents as a guide for choosing one allocation among the allocations in the support of a randomized selection rule.  

We do not think of global monotonicity as the only sensible definition for monotonicity in the context of fair allocation. In particular, it is realistic to expect selection rules to sometimes be randomized rather than deterministic (for example, when goods are indivisible and there are fewer goods than agents). Hence, we find it informative to also consider a notion of ``individual" monotonicity.
Individual monotonicity can be thought of as monotonicity from the point of view of an individual agent who does not know which selection rule will be used. An optimistic agent may hope that the rule will select the allocation that is most favorable to the agent (among the acceptable allocations), whereas a pessimistic agent might fear that the rule will select the allocation that is least favorable to the agent. A fairness notion is individually monotone if it is considered to be monotone both by  optimistic agents (this is {\em upper monotonicity}) and by pessimistic agents (this is {\em lower monotonicity}).

\begin{definition}
\label{def:monotone}
    A fairness notion is {\em individually monotone} if its acceptable allocations satisfy the following properties. For every allocation instance, every agent $i$, and any two vectors of entitlement $b$ and $b'$, where $b >_i b'$, for which the respective sets of acceptable allocations are nonempty, the following holds. 
    \begin{itemize}
        \item {\em Upper monotonicity:} There is an allocation $A = (A_1, \ldots, A_n)$ that is acceptable with respect to $b$ such that for every allocation $A' = (A'_1, \ldots, A'_n)$ that is acceptable with respect to $b'$ it holds that $v_i(A_i) \ge v_i(A'_i)$.
        \item {\em Lower monotonicity:} There is an allocation $A' = (A'_1, \ldots, A'_n)$ that is acceptable with respect to $b'$ such that for every allocation $A = (A_1, \ldots, A_n)$ that is acceptable with respect to $b$ it holds that $v_i(A_i) \ge v_i(A'_i)$. (Remark: we shall often desire a stronger version of lower monotonicity, in which $A'$ referred to above has to be in the Pareto front of the set of allocations acceptable under $b'$.)
    \end{itemize}
\end{definition}

We alert the reader that being individually monotone (Definition~\ref{def:monotone}) does not imply being globally monotone (Definition~\ref{def:global}), and likewise, being globally monotone does not imply being individually monotone.
As we shall see, none of the fairness notions considered in this paper is individually monotone. However, there are different levels of violations of individual monotonicity. In particular, some of them imply also violation of global monotonicity, and some do not.
One way of capturing the extent to which monotonicity is violated (in an obvious way, without the need to go into complicated reasoning about multiple vectors of entitlements), is through what we refer to as {\em paradoxes}. In our setting, a paradox is a situation in which it is in the best interest of an agent (who wishes to maximize the value of the bundle that she receives) to distribute some of her entitlement to other agents.

Our strong form of a paradox involves the notion of an {\em inversion}.

\begin{definition}
    \label{def:inversion}
    A fairness notion is said to suffer from an {\em inversion} if there is an  allocation instance and two vectors of entitlement $b$ and $b'$, where $b >_i b'$ for some $i$, for which the respective sets of acceptable allocations are nonempty and the following holds. 
    {For every allocation $A' = (A'_1, \ldots, A'_n)$ that is acceptable with respect to $b'$ and every allocation $A = (A_1, \ldots, A_n)$ that is acceptable with respect to $b$ it holds that $v_i(A'_i) > v_i(A_i)$.}
\end{definition}

If a fairness notion suffers from an inversion as in Definition~\ref{def:inversion}, we say that it suffers from the {\em inversion paradox}. A fairness notion that suffers from the inversion paradox cannot be globally monotone, and cannot be individually monotone (and moreover, is neither upper monotone nor lower monotone). 



Our weak form of a paradox involves the notion of {\em inverse-domination}.

\begin{definition}
    \label{def:inverse-domination}
    A fairness notion is said to suffer from {\em inverse-domination} if there is an  allocation instance and two vectors of entitlement $b$ and $b'$, where $b >_i b'$ for some $i$, for which the respective sets of acceptable allocations are nonempty and satisfy the following. There is some value $t$ such that all allocations that are acceptable under $b$ give $i$ value at most $t$, all allocations that are acceptable under $b'$ give $i$ value at least $t$, and at least one of the following two possibilities hold:
    \begin{itemize}
        \item Some allocation $A'$ acceptable under $b'$ gives $i$ value strictly more than $t$. (In this case the fairness notion cannot be upper monotone, but might be lower monotone.) 
        \item Some allocation $A$ acceptable under $b$ gives $i$ value strictly less than $t$. (In this case the fairness notion cannot be lower monotone, but might be upper monotone.) (Remark: we shall often desire a stronger version of this inverse domination, in which $A$ referred to above has to be in the Pareto front of the set of allocations acceptable under $b$.)
    \end{itemize}    
\end{definition}

If a fairness notion suffers from inverse-domination as in Definition~\ref{def:inverse-domination}, we say that it suffers from the {\em incentive paradox}. Agent $i$ has an incentive to switch from $b$ to $b'$, despite the fact that $b >_i b'$: she will not lose from such a switch, and she might gain.  For every consistent selection rule (that selects an acceptable allocation when acceptable allocations exist), under $b'$  agent $i$ gets a bundle with at least as high value as under $b$, and for some consistent selection rule the value is strictly higher. 

The incentive paradox is weaker than the inversion paradox, and in particular, does not by itself imply that the associated fairness notion is not globally monotone. 

The following proposition summarizes the relations between various notions associated with monotonicity.

\begin{proposition}
    \label{pro:monotoneProperties}
    The following relations hold between monotonicity properties and paradoxes.
    \begin{itemize}
    \item An individually monotone fairness notion need not be globally monotone and a globally monotone fairness notion need not be individually monotone. 
    \item If a fairness notion suffers from the inversion paradox then it also suffers from the incentive paradox, but suffering from the incentive paradox does not imply suffering from the inversion paradox.
    \item If a fairness notion is individually monotone then it is both upper monotone and lower monotone (by definition), and it does not suffer from any of the two paradoxes. 
    \item If a fairness notion is either upper monotone or lower monotone then it does not suffer from the inversion paradox, but it might suffer from the incentive paradox. 
    \item If a fairness notion suffers from the incentive paradox then it is not individually monotone, but it might be either upper monotone or lower monotone, and it might be globally monotone. 
    \item If a fairness notion suffers from the inversion paradox then it is not globally monotone, not upper monotone and not lower monotone.
    \end{itemize}
\end{proposition}

\subsection{Attitude towards social welfare}
\label{sec:attituteDef}

Another dimension by which we compare fairness notions is that of their attitude towards social welfare. We address this issue in two different ways.

One aspect of respecting social welfare is that of outputting an allocation that is Pareto optimal (see Definition~\ref{def:pareto}). Allocations that can be Pareto-improved can be viewed as wasting welfare. Here we consider three disjoint classes of fairness notions. (They do not cover all possibilities, but suffice for our purpose.)

\begin{itemize}
\item {\em Pareto}. Every acceptable allocation is Pareto optimal.
    \item {\em pro-Pareto}. Every allocation that Pareto dominates an acceptable allocation is acceptable.
    \item {\em non-Pareto}. Some instances have acceptable allocations, but do not have Pareto optimal allocations that are acceptable.  
\end{itemize}

In Section~\ref{sec:pareto} we present the classification of our six fairness notions into these three classes. For CE, our classification corrects inaccurate claims that were made in some of the previous work. 

Another aspect of respecting social welfare is studied most conveniently in the setting of a divisible homogeneous good (and arbitrary valuation functions). In this setting, the proportional allocation, giving each agent a fraction of the good of size equal to her entitlement, is a natural choice. Indeed for some of the fairness notions, this is an acceptable allocation. However, for some other fairness notion, the proportional allocation is sometimes not acceptable. Here we attempt to classify fairness notions based on their deviations from the proportional allocation. The way we evaluate the deviation is by the {\em weighted Nash Social Welfare} (WNSW) of the allocation, where the WNSW of an allocation $A=(A_1, \ldots, A_n)$ for agents with valuations $v_1, \ldots, v_n$ and entitlements  $b_1, \ldots, b_n$ is $\prod (v_i(A_i))^{b_i}$. For allocation of a homogeneous divisible good,
we introduce the following disjoint classes of fairness notions.

\begin{itemize}
    \item {\em Neutral}. The proportional allocation is acceptable. Moreover, in every acceptable allocation, for every agent $i$, the fraction $A_i$ received satisfies $v_i(A_i) \ge v_i(b_i)$. (Recall that for a divisible homogeneous good, $v_i$ is a function that maps the fraction of the good received to a value. Both $A_i$ and $b_i$ are fractions, and hence valid inputs to $v_i$.) In particular, if $v_i$ is strictly monotone, then $A_i \ge b_i$, and if all valuations are strictly monotone, then the proportional allocation is the only acceptable allocation. Note that for every acceptable allocation, the WNSW is at least as high as that of the proportional allocation.
     \item {\em risky-WNSW}. The proportional allocation is acceptable. There are instances in which additional allocations are acceptable, and moreover, among these other acceptable allocations, some have strictly lower WNSW than the proportional allocation.
    \item {\em pro-WNSW}. There are instances in which the proportional allocation is not acceptable. In every acceptable allocation, the WNSW is at least as high as that of the proportional allocation. 
    \item {\em non-WNSW}. There are instances in which the proportional allocation is not acceptable. Moreover, there are instances in which in every acceptable allocation, the WNSW is strictly lower than that of the proportional allocation. 
\end{itemize}

The notion of MWNSW is pro-WNSW by definition. We observe that some of the other fairness notions are non-WNSW. In other words, on some allocation instances, they dictate to replace the proportional allocation by an allocation of lower WNSW. 

\subsection{Incentives}

In our paper we assume that the entitlements of agents and their valuation functions are known to the allocation mechanism. In practice, one may expect the entitlements to be known and verifiable by the mechanism (e.g., there may be a legal document stating the entitlements), but the valuation functions of agents are typically private information of the agents, and in most settings the mechanism cannot verify that the valuations reported to the mechanism are indeed the true valuations of the agents. This opens the possibility of manipulations on behalf of strategic agents, who might report incorrect valuation functions in hope of receiving a bundle of higher value. To address such concerns, it is desirable that an allocation mechanism be {\em incentive compatible}, in the sense that even strategic agents judge that it is in their best interest to report their true valuation functions.

Incentive compatibility is an issue that may be thought of as being independent of that of fairness. 
Nevertheless, for a given fairness notion, it is desirable that there will be incentive compatible allocation mechanisms that are consistent with the fairness notion.

For strong notions of incentive compatability, such as having dominant strategies, results such as those in~\cite{ABCM17} suggest that there is not much hope of finding allocation mechanisms that are both incentive compatible and fair. There are weaker notions of incentive compatibility that some fairness notions do enjoy. One such example, which is applicable to share based fairness notions, is that of a share being {\em self maximizing}~\cite{BF22}. This property is enjoyed by APS, but not by Prop and WMMS (see Section~\ref{sec:self-maximizing}). 

In this work, we shall not consider incentive properties of our fairness notions. However, we do wish to point out that for a divisible homogeneous good, the mechanism that outputs the proportional allocation has excellent incentive properties, because this allocation depends only on the publicly known entitlements, so agents cannot manipulate the outcome.

\subsection{Related work}
\label{sec:related}

There is vast literature on fair allocation. General references includes books such as~\cite{brams1996fair} and~\cite{moulin2004fair} and surveys such as~\cite{amanatidis2023fair}.

In our work we consider several properties of fairness notions, which include feasibility, monotonicity and welfare. We discuss here some past work that is most relevant in this context.

Some of the literature on fair allocation, often referred to as cake cutting, involves establishing feasibility (with a finite number of cuts) for fairness notions when allocating a divisible non-homogeneous good to agents that have additive valuations. For Prop and equal entitlements feasibility was established in~\cite{Steinhaus48}, whereas for WEF and arbitrary entitlements feasibility was established in~\cite{RW97}. For the case of indivisible goods, many fairness notions are not feasible. A central result in this aspect is that the MMS is not feasible even if agents are restricted to additive valuations~\cite{KPW18}. The results that we present for feasibility were all either previously known, or are easy to establish. 

Our study of monotonicity is motivated by the question of whether for a given fairness notion, having higher entitlement is better than having lower entitlement. 
Monotonicity was studied from several angels in the past, some of them not related to this motivating question. 

{\em Resource monotonicity} is a constraint on how an allocation may change if the set of items changes. It postulates that if more items are added, the value received by any agent should not decrease. Though this is a natural property to ask for, it is not motivated by our question above, and we do not address it in this work. 

{\em Population monotonicity}~\cite{Chun86} is a notion which is most natural in the case of equal entitlements. It postulates that if an additional agent is added, the value received by any of the original agents should not increase. Equivalently, if an agent is removed,  the value received by any of the remaining agents should not decrease. For additive valuations, the fairness notion of Maximum Nash Social Welfare (MNSW, we study its extension MWNSW to the case of arbitrary entitlements) was shown to satisfy population monotonicity if items are divisible~\cite{SS19}, but not to satisfy it if items are indivisible~\cite{CSS21}. Our notions of monotonicity differ from population monotonicity. We require monoticity to hold only between vectors of entitlements that satisfy $b >_i b'$ (the entitlement increases only for one agent $i$). Population monotonicity considers pairs of entitlement vectors, such that one of them $b'$ contains $n+1$ agents and the other $b$ contains $n$ agents. Hence $n$ agents increase their entitlements from $\frac{1}{n+1}$ to $\frac{1}{n}$, and there is no single agent $i$ for which $b >_i b'$ (unless $n=1$, an uninteresting case). 

{\em Weight monotonicity}~\cite{CSS21} postulates that if the entitlement of a single agent $i$ increases, whereas for all other agents the entitlements are scaled by the same multiplicative factor $\rho < 1$ (so that entitlements still sum up to~1), then the value received by agent $i$ should not decrease. Like our notions of monotonicity, weight monotonicity considers pairs of entitlement vectors that satisfy $b >_i b'$. However, the difference is that weight monotinicy also dictates in what fashion the entitlements of other agents decrease, and hence is easier to satisfy then our notions of monotonicity. A striking example is MWNSW, for which~\cite{CSS21} shows that weight monotonicity holds, whereas we show that it is neither globally nor individually monotone.

We refer to various violations of monotonicity as {\em paradoxes}, and specifically introduce the inversion paradox and the incentive paradox. Other paradoxes were documented in practice, specifically, in allocating seats to states in the US House of Representatives. The so called {\em Alabama paradox} (discovered in 1880) demonstrated that the allocation method used at the time violated resource monotonicity (adding seats to the house had the potential of reducing the number of seats allocated to the state of Alabama).  There is also a paradox referred to as the {\em population paradox}, though technically, it does not refer to population monotonicity. In this paradox, the term ``population" refers to the population of a state, which translates in our language to its entitlement. In our language, the population paradox refers to a situation in which the entitlements of two of the agents increase, one of them loses value, and the one to lose value is the one whose entitlement increased by the larger multiplicative factor. This paradox does not involve vectors of entitlements satisfying $b >_i b'$, because two agents enjoy increased entitlements. 

In our work we consider two aspects related to welfare. One is the question of whether the underlying fairness notions are in conflict with selecting allocations that are Pareto optimal. The issue of Pareto optimality was extensively studied in the past, and the results presented in our work were either previously known, or are easy to establish. The other aspect, we we discuss only in the context of a divisible homogeneous good, concerns how the use of a given fairness notion affects weighted Nash Social Welfare, compared to the natural proportional allocation. We are not aware of previous studies that address this specific question.

Our work offers a classification of fairness notions based on their properties. In this respect, it is part of a framework of an axiomatic approach for fair allocation (see~\cite{MoulinBook} for axiomatic approaches in somewhat more general settings). Examples of other classifications (based on other properties) include~\cite{Moulin92} for some settings with monetary transfers, and~\cite{SS19} for some settings without monetary transfers. Earlier classifications typically considered {\em allocation rules} (for which the set of acceptable allocations is always non-empty, in the settings considered), whereas we consider fairness notions (for which the set of acceptable allocations might be empty).

\subsection{Outline}

The fairness notions that are considered in this work are defined and briefly discussed in Section~\ref{sec:notions}. Afterwards, we present the classifications of these notions with respect to aspects considered in this paper. Section~\ref{sec:feasibility} presents the classification with respect to feasibility.  Section~\ref{sec:monotone} presents the classification with respect to monotonicity.  Section~\ref{sec:welfare} presents the classification with respect to attitude towards welfare. After the presentation of the classifications, in Section~\ref{sec:discussion} we attempt to draw some conclusions from the classification.

\section{The fairness notions}
\label{sec:notions}

The fairness notions addressed in this work all share the following two properties.

\begin{itemize}
    \item For every allocation instance, the set of acceptable allocations depends only on the entitlements of the agents and on their valuation functions.
    \item The fairness notions are scale invariant: scaling the valuation of an agent by a multiplicative constant does not change the set of acceptable allocations.
\end{itemize}

It may be instructive to think of the following example. Suppose that a father owns a violin that he wishes to give to one of his two twin daughters. He loves both daughters just the same. One of the daughters is a gifted violin player and can play the violin to her own enjoyment, whereas the other cannot play the violin at all. We may model the situation as that of allocating an indivisible item (the violin) to two agents (the daughters) of equal entitlement (the father loves both daughters equally), where one daughter has high value for the violin (she can play it), whereas the other has low value (she cannot play it). Under this modeling, none of the fairness notions considered in this paper postulates that one daughter should have any advantage over the other in receiving the violin, because the fairness notions are scale invariant. However, if we model the situation as that of unequal entitlements (e.g., the daughter who plays the violin has higher entitlement, due to the hard work that she put into learning to play the violin), then scale invariant notions may give an advantage to the daughter who plays the violin.



Some of the fairness notions that we consider are typically defined in settings with a set $\items$ of $m$ indivisible items. To extend such definitions to the setting of a divisible good, 
we view the divisible good as a collection of $m$ identical indivisible goods. Here, $m$ is assumed to be huge (its size may depend on the entitlements and valuations of the agents), so that for each fixed fraction $f_i$ of interest, allocating a fraction $f_i$ of the indivisible good is equivalent to allocating $f_i m$ indivisible items
(we shall not be interested in dealing with cases in which $f_i$ is irrational). 

For each fairness notion we present both its formal definition, and an intuitive way to think about it. This intuition is presented as a plausible explanation of why no agent has a justifiable reason to complain, when the allocation is acceptable according to the fairness notion. These explanations should not be interpreted as implying that when the set of acceptable allocations contains more than one allocation, all acceptable allocations are equally desirable, or equally fair. A selection rule may well prefer one of them over the others.


\subsection{Share based fairness notions}

In our presentation we shall distinguish between two notions of shares, referring to one of them as {\em proper} shares (they coincide with the definition of shares as given for example in~\cite{BF22}), and the other as {\em general} shares (which allows for wider class of shares). 

A {\em proper} share function $s$ maps two arguments, the entitlement $b_i$ of an agent $i$ and her valuation function $v_i$, into a non-negative value $s(v_i, b_i)$. For allocation of goods, the share function $s$ is required to be nondecreasing in $b_i$. An allocation is acceptable according to $s$ if every agent $i$ gets a bundle $A_i$ of value at least $v_i(A_i) \ge s(v_i, b_i)$. We present some notions of proper shares that are commonly used.

\begin{definition}
\label{def:prop}
    The {\em proportional share} (Prop) is $Prop(v_i,b_i) = b_i \cdot v_i(\items)$.
\end{definition}

For an allocation acceptable under Prop, an agent with entitlement $b_i$ has no justifiable reason to complain, because the fraction of total value that she received is at least her fraction of the total entitlement. 
\medskip

The maximin share (MMS) was originally defined only for instances in which agents have equal entitlement~\cite{Budish11}, though the definition can be applied to any agent that has entitlement of the form $b_i = \frac{1}{k}$, where $k$ is an integer (regardless of entitlements of other agents).

\begin{definition}
\label{def:MMS}
    The {\em maximin share} (MMS) is defined only if $b_i = \frac{1}{k}$ for some integer $k$. Its value is the maximum over all partitions $(B_1, \ldots, B_k)$ of the items into $k$ bundles, of the minimum value $\min_j [v_i(B_j)]$ of a bundle in the partition. 
\end{definition}

For an allocation acceptable under MMS, an agent with entitlement $\frac{1}{k}$ has no justifiable reason to complain about her bundle $A_i$, because if all other agents are identical to her (have the same entitlement and same valuation function), then it is unavoidable that some agent would not get a bundle of value higher than that of $A_i$.
\medskip

An extension of MMS to arbitrary entitlements was proposed in~\cite{BNT21}. It is referred to as the $\ell$ out of $d$ share in~\cite{SegalHalevi20}, and as the pessimistic share in~\cite{BEF21APS}.

\begin{definition}
\label{def:pess}
    The {\em pessimistic share} (Pess) of agent $i$ is the maximum 
    over all choices of integers $d$ and partitions $(B_1, \ldots, B_d)$ of the items into $d$ bundles, of the minimum value $\min_{\{J : |J|=\ell\}} [v_i(B_{\{J\}})]$. Here, $J$ is a set of $\ell$ indices from $\{1, \ldots, d\}$), where $\ell$ is the largest integer satisfying $\frac{\ell}{d} \le b_i$, and $B_{\{J\}} = \cup_{j\in J} B_j$. 
\end{definition}

The definition of Pess is presented here only for completeness. It will not be one of the six fairness notions that we compare, because to a large extent its properties are similar to those of the APS, a fairness notion that will be included among the six.
\medskip

The {\em anyprice share} (APS) has two equivalent definitions~\cite{BEF21APS}. One is a min-max definition involving prices. 

\begin{definition}
\label{def:APS1}
    The {\em anyprice share} (APS) of agent $i$ is the minimum over all admissible price vectors $(p_1, \ldots, p_m)$ (admissible in the sense of being non-negative and summing up to~1) of the maximum value of a set affordable with budget $b_i$, namely, of $\max_{\{S \mid \sum_{j\in S} p_j \le b_i\}} [v_i(S)]$. 
\end{definition}

For an allocation acceptable under APS, an agent $i$ who received bundle $A_i$ has no justifiable reason to complain, because under certain item prices, any bundle with higher $v_i$ value than $A_i$ would cost strictly more than a $b_i$ fraction of total cost of all items, whereas agent $i$ is entitled to only a $b_i$ fraction of the total cost. 

The other definition of the APS is a max-min definition that may be thought of as a fractional version of the definition of the MMS.

\begin{definition}
\label{def:APS2}
    A fractional partition of $\items$ is a collection $(B_1, B_2, \ldots)$ of subsets of $\items$ and associated nonnegative weights $(\lambda_1, \lambda_2 \ldots)$ such that $\sum_j \lambda_j = 1$. The fractional partition is $c$-bounded if for every item $e$, $\sum_{j \mid e \in B_j} \lambda_j \le c$. The {\em anyprice share} (APS) of agent $i$ is the maximum value $t$ such that there is a $b_i$-bounded fractional partition of $\items$ that contains only bundles of $v_i$ value at least $t$. 
\end{definition}
\medskip

A different extension of MMS to arbitrary entitlements was proposed in~\cite{FGHLPSSY19}, and is referred to as the {\em weighted maximin share} (WMMS). It is a {\em general} share rather than a {\em proper} share, because its value depends not only on the entitlement and valuation of the given agent, but also on the entitlements of other agents. 

\begin{definition}
\label{def:WMMS}
    Given the vector $(b_1, \ldots, b_n)$ of entitlements, the {\em weighted maximin share} (WMMS) of agent $i$ is the maximum over all partitions $(A_1, \ldots, A_n)$ of the items into $n$ bundles, of the minimum value $\min_{j} [\frac{b_i}{b_j}v_i(A_j)]$. 
\end{definition}

Here is a plausible justification for why allocations acceptable under WMMS are fair.  Consider agent $i$ with valuation $v_i$ and entitlement $b_i$ who receives bundle $A_i$. Her value per unit entitlement is $\frac{v_i(A_i)}{b_i}$. This value is not too low, by the following mental experiment. Suppose all other agents keep their entitlements, but have the same valuation function $v_i$ that agent $i$ has. Then it is unavoidable that at least one agent would get a bundle for which the value per unit entitlement is no larger than $\frac{v_i(A_i)}{b_i}$.
\medskip

We state some known relations between the various notions of shares. (For the proof of item~3, see~\cite{BEF21APS}.)

\begin{proposition}
\label{pro:relations}
    The following relations hold between share values:
    \begin{enumerate}
        \item  When the entitlement is $\frac{1}{k}$ for integer $k$ (which holds in particular in the equal entitlement case), the values of the the pessimistic share and the WMMS are both equal to the MMS.
        \item When valuations are additive, then the proportional share has value at least as large as any of the other shares defined above. However, when valuations are not additive, the value of the proportional share might be lower than the values of any of the other shares defined above. 
        \item The value of the APS is always at least as large as the pessimistic share. Moreover, it is sometimes strictly larger, even for instances with equal entitlements (for which the pessimistic share equals the MMS) and additive valuations.
    \end{enumerate}
\end{proposition}

We remark that in contrast to item~3 in Proposition~\ref{pro:relations}, for the case of a divisible homogeneous good, the APS and the pessimistic share have the same value. 

Intuitively, share based fairness notions model agents as self-centered. To agree that an allocation is acceptable, an agent inspects only the value of the bundle that she herself receives, and is indifferent towards the values of bundles received by other agents.


\subsection{Comparison based fairness notions}

Here we consider fairness notions in which an agent compares the bundle that she receives with bundles received by other agents. Such share notions may be referred to as {\em comparison based}, though standard terminology often uses the term {\em envy}.

\begin{definition}
\label{def:EF}
    An allocation among agents of equal entitlement is {\em envy-free} (EF) if no agent strictly prefers a bundle received by a different agent over her own bundle.
\end{definition}

The empty allocation that does not allocate any item is EF in a trivial sense. To avoid such trivialities, one often adds a requirement that all items are allocated. In the setting of allocating a divisible good, we require that the whole good is allocated. (For share based fairness notions, this requirement holds without loss of generality, because every acceptable partial allocation (acceptable in the sense that every agent gets at least her share value) can be completed to an acceptable full allocation.)

We note the following relation between EF allocations and allocations that are acceptable under various share notions.

\begin{proposition}
\label{pro:EF}
    For agents with additive valuations, EF allocations give every agent at least her proportional share (and hence, also at least her MMS). However, for every $\epsilon$, there are instances with agents with non-additive valuations (and in particular, submodular valuations) that have EF allocations in which every agent gets at most an $\varepsilon$-fraction of her MMS.
\end{proposition}

\begin{proof}
    We just present a negative example illustrating the second statement of the proposition. Let $n > \frac{1}{\varepsilon}$, and suppose that there are $n^2$ items, partitioned into $n$ equal sized groups of items. All agents have the same submodular valuation, where the value of a set $S$ of items is the number of groups that $S$ intersects. The MMS of each agent is $n$, with an MMS partition being $n$ disjoint bundles, where each bundle has one item from each group.  The allocation that gives each agent one group of items allocates all items. In this allocation, each agent gets value~1. This allocation is EF, but gives each agent less that $\varepsilon$-MMS. 
\end{proof}

Due to examples such as the one presented in the proof of Proposition~\ref{pro:EF}, one sometimes places an additional requirement on EF allocations, that the allocation will be Pareto optimal. We do not make such a requirement in this manuscript (though we do require that all items be allocated). 

We present here an extension of envy-freeness proposed in~\cite{RP98}, to settings with unequal entitlements.

\begin{definition}
\label{def:WEF}
    An allocation $(A_1, \ldots, A_n)$ to agents of arbitrary entitlements (with entitlements $(b_1, \ldots, b_n)$ and valuations $(v_1, \ldots, v_n)$) is {\em weighted envy-free} (WEF) if no agent $i$ envies another agent $j$, after scaling the value of the bundle received by $j$ so as to reflect the ratio of entitlements between $i$ and $j$. Formally, for every two agents $i$ and $j$, it holds that $v_i(A_i) \ge \frac{b_i}{b_j}v_i(A_j)$. 
\end{definition}

Intuitively, envy based fairness notions model agents as competitors. To agree that an allocation is fair, an agents needs to inspect the value of the bundles received by all agents, and convince herself that no other agent got a better deal than she did. An agent is not discontent with a bundle of low value, as long as all other agents suffer as well. 

\begin{remark}
    For WEF, it may be natural to introduce monotonicity properties that are different from those presented in Section~\ref{sec:monotone}, and take into account the change in values of bundles received by all agents (because the desirability of a bundle to an agent driven by envy depends on which bundles the other agents receive). This is addressed in Section~\ref{sec:disWEF}. 
\end{remark}


\subsection{Optimization based fairness notion}

One may set up an objective function, and declare the allocation that maximizes this objective function as the one that should be chosen. (In case of ties, any of the maximizing allocations is acceptable.) Optimization based fairness notions attempt to select an objective function that is aligned well with intuitive notions of fairness, and to select an allocation that optimizes this objective. Given an allocation instance, the objective function induces a total order (not necessarily strict -- it might have ties) over allocations, and one selects a maximal element in this total order. 

In the equal entitlement case, the {\em min order} is sometimes used as the basis for optimization based fairness notions: allocations are ordered in the order of the value received by the least happy agent. Allocations that maximize the respective objective are referred to as max-min allocations~\cite{BD05}. A natural refinement of this order is the {\em leximin order}, which in cases of ties in the min order considers the value received by the next to least happy agent, and if these values are also tied moves to the next agent, and so on. 

The above orders are not scale invariant: scaling the valuation function of an agent by a multiplicative factor may change the order. Hence (in our opinion) these orders should be used only in situations in which the values for all agents can be measured in the same universal scale (such as in US dollars). Alternatively, one can perform some canonical scaling of valuations before using these orders. This canonical scaling can be based on a share based fairness notion. For example, one can scale each valuation function such that the maximin share of the respective agent is~1. Moreover, this principle can be used to easily extend the use of leximin order to settings of unequal entitlements, by scaling the valuation function of each agent so that her APS is~1.

In this manuscript we take the view that leximin order should be used only after canonical scaling of valuations, and furthermore, that this scaling should be based on a share based notion, such as the APS. Under this view, in cases were acceptable APS-allocations exist, the only role of the leximin part is to serve as a selection rule among the acceptable allocations. That is, we view share based notions as providing fairness constraints that determine which allocations are acceptable, and leximin as a principle for selecting one allocation among the acceptable ones. 

Another way of achieving scale invariance in the order is to select an order that is scale invariant to begin with. This is accomplished by considering the Nash social welfare (NSW).

\begin{definition}
\label{def:NSW}
    For agets with equal entitlement, the {\em Nash Social Welfare} (NSW) of an allocation $(A_1, \ldots, A_n)$ is $\left(\prod_{i=1}^n v_i(A_i) \right)^{1/n}$, the geometric mean of the values received by the agents. 
\end{definition}

Allocations that maximize NSW are often advocated as either being fair, or at least as offering a good balance between the total welfare of the population and fairness towards individual agents.  In certain cases, they are known to imply other fairness properties~\cite{CKMPSW19}.

A common extension of NSW to the case of unequal entitlements is the {\em Weighted Nash Social Welfare} (WNSW)~\cite{CISZ21}.

\begin{definition}
\label{def:WNSW}
For agents of possibly unequal entitlement $(b_1, \ldots, b_n)$, the {\em Weighted Nash Social Welfare} (WNSW) is $\prod_{i=1}^n (v_i(A_i))^{b_i}$.   
\end{definition}

The Maximum Weighted Nash Social Welfare (MWNSW) fairness notion postulated that the allocation rule should select an allocation that maximizes WNSW. In general, ties can be broken arbitrarily, though in the special case in which the MWNSW is~0, a more careful tie braking rule is used. Specifically, one selects a set of agents of maximum size such that there is an allocation that gives each of them positive value, and maximizes WNSW for this set of agents.

Intuitively, accepting maximum Nash Social Welfare as a fairness notion models agents as being cooperative. To agree that an allocation is fair, an agents needs to inspect the value of the bundles received by all agents, and to know the valuation functions of all agents. An agent is not discontent with a bundle of low value, as long as other agents gain from this. Agents are willing to lose a little in the value of bundles that they receive, if other agents gain a lot.


\subsection{Equilibrium based fairness notions}

Here we consider a fairness notion that draws its inspiration from economic theories concerning equilibrium market prices~\cite{Budish11,BNT21}.

\begin{definition}
\label{def:CE}
    Given an allocation instance with $n$ agents with valuations $(v_1, \ldots, v_n)$ and entitlements $(b_1, \ldots, b_n)$, an allocation $(A_1, \ldots, A_n)$ is a {\em competitive equilibrium} (CE) if there are non-negative prices $(p_1, \ldots, p_m)$ to the items such that for every agent $i$, the bundle $A_i$ is affordable (its total price is at most $b_i$), and no bundle affordable to $i$ has higher value (according to $v_i$). For divisible items, we assume that an $\alpha$-fraction of an item costs an $\alpha$-fraction of the total cost of the item. 
\end{definition}

Definition~\ref{def:CE} does not require that there are prices in which every agent $i$ spends her budget $b_i$. For example, in an instance with two agents of unequal entitlement and a single indivisible good, allocating the good to the agent with higher entitlement is a CE (e.g., by pricing the good at $\frac{1}{2}$), but the lower entitlement agent does not get to spent any part of her budget. 


\begin{remark}
\label{rem:PCE}
    There is a variation on the definition of CE, introduced in~\cite{MasColell}, and later referred to as parsimonious CE (PCE), in which among all affordable bundles of maximum value, the bundle allocated to the agent is one of lowest price. This variation is motivated by issues of Pareto optimality. See Section~\ref{sec:disCE}. 
\end{remark} 

Competitive equilibrium implies some of the other fairness notions that we encountered. The proof of the following proposition follows almost immediately from the respective definitions, and is omitted. 

\begin{proposition}
\label{pro:CEimplies}
Every CE allocation gives each agent at least her APS.
    In the equal entitlement case, every CE allocation is envy-free. 
\end{proposition}

Intuitively, competitive equilibrium models agents as respecting the wishes of other agents, but indifferent to the welfare of other agents. The equilibrium prices are postulated to faithfully represent the objective value for the items, and the agent should get her fair share (her part of the general entitlement) with respect to these prices. The agent is respectful of the wishes of other agents, in the sense that the prices reflect these wishes. The agent is indifferent to what other agents actually get, in the sense that based on the given prices, the agent selects the best bundle that she can get.


\section{Feasibility for fairness notions}
\label{sec:feasibility}

In this section we discuss feasibility of the six fairness notions that are the focus of our work, in several different contexts.

For the case of indivisible items, none of our share based fairness notions (Prop, WMMS, APS) is feasible, not even if valuations are additive and agent have equal entitlements. This was shown for the MMS in~\cite{KPW18}, and follows for the shares considered here because for such instances, their values are at least as large as the MMS (see Proposition~\ref{pro:relations}).
The fairness notions of EF and CE are also clearly not feasible, e.g., when there are fewer goods than agents, and agents have equal entitlements. In contrast, MWNSW is feasible for every class of valuation functions, because having a only a finite number of possible allocations implies that a maximal allocation (maximizing WNSW) exists.

For a divisible homogeneous good, both APS and CE are feasible for every class of valuation functions, as the proportional allocation is acceptable under these notions. As to MWNSW, it too is feasible, except for what we view as a technicality. The technical issue is the following: if the valuation functions are not continuous, it might be the case that no maximal allocation exists. For example, consider one divisible good and two agents of equal entitlement. The first agent receives value~1 if she gets strictly more than half the good and value~0 otherwise. The second agent gets value equal to the fraction of the good that she receives. Then as $\epsilon$ decreases, the Nash social welfare of the allocation $(\frac{1}{2} + \epsilon, \frac{1}{2} -  \epsilon)$ increases, but for $\epsilon=0$ the NSW drops to~0. We believe that such technical issues are more a matter of mathematical modelling and not an issue of real practical significance (for example, these issues do not arise if the divisible homogenuous good is replaced by any finite number $m$ of identical indivisible items), and hence we ignore this issue and view MWNSW as a notion that is always feasible.

Continuing with a divisible homogeneous good, Prop is feasible when valuations are linear or concave (the proportional allocation is acceptable), but not feasible if all valuations are convex, and at least one of them is strictly convex. This is because to get her proportional share $b_i \cdot v_i(\items)$, an agent with a strictly convex valuation (such as $v_i(x) = x^2$) must get strictly more than a $b_i$ fraction of the good.

As to WMMS and WEF, neither one of them is feasible, not even if all valuations are concave. This is demonstrated in the following proposition.

\begin{proposition}
\label{pro:infeasible}
        Consider an instance of allocating a homogeneous divisible good in which there are two agents with entitlements $b_1 = \frac{1}{3}$ and $b_2 = \frac{2}{3}$ and valuations $v_1(x) = x$ and $v_2(x) = \sqrt{x}$. Then:
        \begin{itemize}
            \item In every allocation some agent does not get her WMMS.
            \item There is no WEF allocation.
        \end{itemize} 
\end{proposition}

\begin{proof}
    The WMMS of agent~1 is $\frac{1}{3}$, whereas that of agent~2 is $\frac{4}{5}$ (with entitlements $\frac{1}{3}, \frac{2}{3}$ and valuation $v_2(x) = \sqrt{x}$ the WMMS partition is $(\frac{1}{5},\frac{4}{5})$). As these values sum to above~1, there is no WMMS allocation. Likewise, an allocation $(a, 1-a)$ is WEF only if $a \ge \frac{1}{3}$ and $1 - a \ge \frac{4}{5}$, and so there is no WEF allocation. 
\end{proof}

For WMMS there is an interesting case in which it is always feasible (both for divisible items and indivisible ones), and that is the case of an external valuation function (that is, all agents have the same valuation function). In that case, all agents compute the value of their WMMS share using the same WMMS partition, and this WMMS partition can serve as an acceptable allocation. If the item is divisible and the external valuation function is continuous, then a WEF allocation also exists. 

We summarize the results concerning feasibility in Table~\ref{tab:feasibility}. In general, in every table that we present, the fairness notions are grouped together based on their performance with respect to the properties considered in the table.  


\begin{table}[hbt!]
        \centering
\begin{tabular}{ |c||c|c| } 
 \hline
 \multicolumn{3}{|c|}{Feasibility} \\
  \hline
  Fairness notion & Homogeneous Divisible Good & Indivisible Goods \\ 
 \hline
 \hline 
 MWNSW & yes$^*$ & yes \\  
 \hline
 APS & yes & no \\
 CE & yes & no \\ 
 \hline
  Prop & no (yes for concave $v_i$'s) & no \\
 \hline
 WMMS & no (yes for external $v$) & no (yes for external $v$) \\ 
 \hline
 WEF & no  & no  \\  
 \hline
\end{tabular}
        \caption{Feasibility for fairness notions considered in this work. The negative results for indivisible goods hold even for additive valuations. ($^*$) For a homogeneous divisible good, if valuations are not continuous, the definition of MWNSW requires technical modifications (as the WNSW function need not have a maximum). See main text.}
        \label{tab:feasibility}
        \end{table}

\section{Monotonicity}
\label{sec:monotone}

In this section we discuss the monotonicity properties of the fairness notions under consideration in this paper. We start by establishing that fairness notions that are based on proper shares are immune to the inversion paradox.

\begin{proposition}
\label{pro:upperMonotone}
    Every fairness notion based on a proper share is upper monotone (for any class of valuations). Consequently, it does not suffer from the inversion paradox.
\end{proposition}

\begin{proof}
Recall that every proper share function $s = s(v_i,b_i)$ is non-decreasing in the entitlement $b_i$. Consider vectors of entitlement $b$ and $b'$, where $b >_i b'$. Consider any allocation $A'$ that is feasible for $b'$. If $v_i(A'_i) \ge s(v_i,b_i)$, then $A'$ is also feasible for $b$ (because for every $j \not= i$, $b_j \le b'_j$) and can serve as $A$. If $v_i(A'_i) < s(v_i,b_i)$, then in any allocation $A$ that is feasible for $b$, $v_i(A_i) > v_i(A'_i)$.
\end{proof}

In our remaining treatment of monotonicity, we consider separately the case of indivisible goods, and that of a divisible homogeneous good.

\subsection{Indivisible goods}

Recall that Prop and APS are upper monotone, and hence do not suffer from the inversion paradox.
We now turn attention to the incentive paradox.

\begin{proposition}
\label{pro:PropIncentive}
    Prop suffers from the incentive paradox, even when valuations are additive.
\end{proposition}

\begin{proof}
    Consider three agents with additive valuations, and three items $\{e_1,e_2,e_3\}$. The valuations are presented in Table~\ref{tab:PropIncentive}:

    \begin{table}[hbt!]
        \centering
        \begin{tabular}{c|ccc}
               & $e_1$ & $e_2$ & $e_3$ \\
               \hline
            $v_1$ & 2 & 1 & 0 \\
                           \hline
            $v_2$ & 0 & 2 & 1\\
                           \hline
            $v_3$ & 3 & 0 & 2\\
        \end{tabular}
        \caption{Incentive paradox for Prop, additive valuations}
        \label{tab:PropIncentive}
    \end{table}
With entitlements $b=(0.2, 0.4, 0.4)$, the proportional shares are $(0.6, 1.2, 2)$. The only acceptable allocation gives item $e_i$ to agent $i$, and agent~1 gets a value of $2$. With entitlements $b'=(0.3, 0.3, 0.4)$, the proportional shares are $(0.9, 0.9, 2)$. The Pareto optimal allocation $\left(\{e_2\},\{e_3\},\{e_1\}\right)$ becomes acceptable and gives agent~1 value of only~1, despite the fact that $b' >_1 b$.
\end{proof}



We do not know whether the APS suffers from the incentive paradox when valuations are additive.  Hence, we enlarge the class of valuations under consideration from additive to {\em budget additive}. A valuation function $v$ is budget additive (BA) is there is a budget $t$ and an additive valuation function $u$ such that for every bundle $S$, $v(S) = \min[t, u(S)]$.

\begin{proposition}
    The APS suffers from the incentive paradox if valuations are {\em budget additive} (BA). The APS is not lower monotone, not even for additive valuations.
\end{proposition}

\begin{proof}
    Consider three agents and five items, $\{e_1, e_2, e_3, e_4, e_5\}$. $v_1$ is budget additive with a budget of~13 (that is, $v_1(S) = \min[13, \sum_{e_j\in S} v_1(e_j)]$), whereas $v_2$ and $v_3$ are additive. The item values are presented in Table~\ref{tab:APSIncentive}:
    \begin{table}[hbt!]
        \centering
        \begin{tabular}{c|ccccc}
               & $e_1$ & $e_2$ & $e_3$ & $e_4$ & $e_5$ \\
               \hline
            $v_1$ & 11 & 2 & 4 & 4 & 4 \\
                           \hline
            $v_2$ & 0 & 2 & 4 & 4 & 4 \\
                           \hline
            $v_3$ & 11 & 0 & 0 & 0 & 0 \\
        \end{tabular}
        \caption{Incentive paradox for APS. $v_1$ is budget additive with a budget of 13.}
        \label{tab:APSIncentive}
    \end{table}
    
    With entitlements $b = (\frac{1}{2}, \frac{1}{3}, \frac{1}{6})$ the APS of the three agents are $(12, 4, 0)$. In every acceptable allocation, agent~2 must get at least one of $e_3, e_4, e_5$, and agent~1 must get $e_1$ and an additional item, for total value of~13 (the maximum possible value for agent~1, given her budget).  With entitlements $b = (0.51, 0.33, 0.16)$ the APS of the three agents are $(12, 2, 0)$. Now the Pareto optimal allocation $(\{e_3, e_4, e_5\}, \{e_2\}, \{e_1\})$ is also acceptable, giving agent~1 a value of~12, despite $b' >_1 b$. 

    The example above also shows that the APS is not lower monotone. In fact, not being lower monotone holds even if $v_1$ is additive rather than budget additive.
\end{proof}

We do not know whether APS is globally monotone, but Prop is not.

\begin{proposition}
    Prop is not globally monotone, not even for additive valuations.
\end{proposition}

\begin{proof}
Consider four items and four agents with valuation functions as in Table~\ref{tab:PropGlobal}.

\begin{table}[hbt!]
        \centering
        \begin{tabular}{c|cccc}
               & $e_1$ & $e_2$ & $e_3$ & $e_4$ \\
               \hline
            $v_1$ & 25 & 29 & 23 & 23 \\
                           \hline
            $v_2$ & 23 & 29 & 25 & 23 \\
                           \hline
            $v_3$ & 23 & 23 & 25 & 29 \\
             \hline
            $v_4$ & 25 & 23 & 23 & 29 \\
        \end{tabular}
        \caption{Prop is not globally monotone, not even for additive valuations.}
        \label{tab:PropGlobal}
        \end{table}

With equal entitlement, $b = (0.25, 0.25, 0.25, 0.25)$, there are only two allocations acceptable under Prop. One is $A^1 = (e_1, e_2, e_3, e_4)$ and the other is $A^2 = (e_2, e_3, e_4, e_1)$. Observe that agents~1 and~3 strictly prefer $A^2$ over $A^1$, whereas agents~2 and~4 strictly prefer $A^1$ over $A^2$.

If the selection rule selects $A^1$, then consider the vector of entitlements $b^1 = (0.24, 0.25, 0.26, 0.25)$. The unique allocation acceptable under Prop is $A^2$. Hence agent~1 strictly prefers the allocation under $b^1$ over that of $b$, despite the fact that $b >_1 b^1$.

If the selection rule selects $A^2$, then consider the vector of entitlements $b^2 = (0.25, 0.24, 0.25, 0.26)$. The unique allocation acceptable under Prop is $A^1$. Hence agent~2 strictly prefers the allocation under $b^2$ over that of $b$, despite the fact that $b >_2 b^2$.
\end{proof}

The remaining fairness notions suffer from the inversion paradox.

\begin{proposition}
\label{pro:notMonotone}
    For allocation of indivisible goods, the following fairness notions suffer from the inversion paradox, even if all agents have additive valuations.
\begin{itemize}
\item Share based notions: WMMS.
    \item Envy based notions: WEF. 
    \item Optimization based notions: MWNSW.
    \item Equilibrium based notions: CE.
\end{itemize}
(Among other things, this implies that these fairness notions are neither upper monotone nor lower monotone.)
\end{proposition}

\begin{proof}
For WMMS consider three agents with additive valuations over four items $(e_1,e_2,e_3,e_4)$. The valuation functions are presented in Table~\ref{tab:WMMSInversion}.

 \begin{table}[hbt!]
        \centering
        \begin{tabular}{c|cccc}
               & $e_1$ & $e_2$ & $e_3$ & $e_4$ \\
               \hline
            $v_1$ & 12 & 20 & 45 & 26 \\
                           \hline
            $v_2$ & 20 & 12 & 45 & 26 \\
                           \hline
            $v_3$ & 10 & 9 & 11 & 11 \\
        \end{tabular}
        \caption{Inversion paradox for WMMS, additive valuations.}
        \label{tab:WMMSInversion}
        \end{table}
        
        With entitlements $0.46, 0.45, 0.09$, the WMMS of agent~1 is 46 (with WMMS partition $(\{e_2, e_4\}, e_3, e_1)$),  the WMMS of agent~2 is 45 (with WMMS partition $(\{e_1, e_4\}, e_3, e_2)$), and the WMMS of agent~2 is $11 \cdot \frac{9}{45} < 3$ (with WMMS partition $(\{e_2, e_4\}, e_3, e_1)$). The only WMMS allocation is $(\{e_2, e_4\}, e_3, e_1)$, giving agent~3 a value of~10. 
        
        With entitlements $0.44, 0.45, 0.11$, the WMMS of agent~1 is $46 \cdot \frac{44}{45} < 45$ (with WMMS partition $(e_3, \{e_2, e_4\}, e_1)$),  the WMMS of agent~2 is 46 (with WMMS partition $(e_3, \{e_1, e_4\}, e_2)$), and the WMMS of agent~2 is $11 \cdot \frac{11}{44} < 3$ (with WMMS partition $(e_3, \{e_2, e_4\}, e_1)$). The only WMMS allocation is $(e_3, \{e_1, e_4\}, e_2)$, giving agent~3 a value of~9. Agent~3 suffers from the inversion paradox.

For WEF, consider three agents with additive valuations. There are five items, with values as in Table~\ref{tab:WEFInversion}.

 \begin{table}[hbt!]
        \centering
        \begin{tabular}{c|ccccc}
               & $e_1$ & $e_2$ & $e_3$ & $e_4$ & $e_5$ \\
               \hline
            $v_1$ & 10 & 10 & 11 & 12 & 37 \\
                           \hline
            $v_2$ & 10 & 10 & 11 & 12 & 37 \\
                           \hline
            $v_3$ & 10 & 10 & 11 & 12 & 57 \\
        \end{tabular}
        \caption{Inversion paradox for WEF, additive valuations.}
        \label{tab:WEFInversion}
    \end{table}

With entitlements $0.12,0.12,0.76$, in the unique WEF allocation,  the first two agents each get an item of value $10$ (having equal entitlement and the same valuation function, they must receive items of the same value), and the third agent gets the remaining items, for a value $80$. (The first two agents do not envy the third, because they value the third bundle only at $60$.) 

With entitlements $0.11, 0.12, 0.77$, in the unique WEF allocation, the first agent gets $11$, the second agent gets $12$, and the third gets $77$. Both agent~1 and agent~3 suffer from the inversion paradox.


For MWNSW, consider three items and three agents with the following additive valuations: 

\begin{table}[hbt!]
        \centering
        \begin{tabular}{c|ccc}
               & $e_1$ & $e_2$ & $e_3$ \\
               \hline
            $v_1$ & 2 & 1 & 0 \\
                           \hline
            $v_2$ & 0 & 1 & 5\\
                           \hline
            $v_3$ & 1 & 0 & 4\\
        \end{tabular}
        \caption{Inversion paradox for MWNSW, additive valuations}
        \label{tab:MWNSWInversion}
    \end{table}

Regardless of the entitlements, there are only two allocations with positive WNSW. In one ($A^1$) each agent $i$ gets item $i$, and the values received by agents are $(2,1,4)$. In the other ($A^2$) each agent $i$ gets item $i + 1$ (and agent~3 gets item~1), and the values received by agents are $(1,5,1)$. Agent~1 gets higher value in $A^1$. With equal entitlement, the MWNSW allocation is indeed $A^1$. Increasing the entitlement of $v_1$ to nearly $\frac{2}{3}$ and decreasing the entitlement of $v_3$ to nearly~0 (making the effect on the WNSW of the allocation of agent~3 negligible), the MWNSW allocation is $A^2$. Agent~1 suffers from the inversion paradox.

For CE, consider three agents and three items $(e_1, e_2, e_3)$. The agents have the following additive valuations:

\begin{table}[hbt!]
        \centering
        \begin{tabular}{c|ccc}
               & $e_1$ & $e_2$ & $e_3$ \\
               \hline
            $v_1$ & 6 & 5 & 4 \\
                           \hline
            $v_2$ & 6 & 4 & 5\\
                           \hline
            $v_3$ & 5 & 4 & 6\\
        \end{tabular}
        \caption{Inversion paradox for CE, additive valuations}
        \label{tab:CEInversion}
    \end{table}


Let $b = (0.36, 0.38, 0.26)$. The minimum entitlement is more than half of the maximum entitlement and the number of items equals the number of agents. Hence, in the unique CE, each agent gets one item, the highest entitlement agent gets her most preferred item, and the second highest entitlement agent gets her most preferred item among those remaining. Thus agent~2 gets $e_1$, agent~1 gets $e_2$ and agent~3 gets $e_3$, for a value of~6.  Consider now $b' = (0.36, 0.34, 0.30)$. In the unique CE, agent~1 gets $e_1$, agent~2 gets $e_3$, and agent~3 gets $e_2$, for a value of only~4, despite the fact that $b' >_3 b$.
\end{proof}

The results concerning monotonicity for allocation of indivisible goods are summarized in Table~\ref{tab:monotonicityIndivisible}. 

\begin{table}[hbt!]
        \centering
\begin{tabular}{ | m{2cm} || m{2cm}| m{2cm} | m{2cm} | m{2cm} |} 
 \hline
 \multicolumn{5}{|c|}{Indivisible Goods} \\
 \hline
  Fairness \newline notion & upper \newline monotone & lower \newline monotone & incentive \newline paradox & inversion \newline paradox\\ 
 \hline
 \hline
 APS & yes & no  & suffers$^*$ & immune \\
 Prop & yes & no   & suffers & immune \\
 \hline
 WMMS & no & no & suffers & suffers\\ 
 WEF & no & no & suffers & suffers\\ 
 MWNSW & no & no & suffers & suffers \\ 
 CE & no & no & suffers & suffers\\ 
 \hline
\end{tabular}
        \caption{None of the fairness notions shown in the table is globally monotone, except perhaps APS, for which this question remains open. $(^*)$ The example showing the incentive paradox for APS involves budget additive valuations.  All other examples for paradoxes hold even with additive valuations.}
        \label{tab:monotonicityIndivisible}
\end{table}

Observe that our results concerning the inversion paradox imply that the notions of WMMA, WEF, MWNSW and CE are not globally monotone (in the sense of Definition~\ref{def:global}), but leave the question open for Prop and APS.

\subsection{Divisible Homogeneous good}

For allocation of a divisible homogeneous  good, some fairness notions exhibit improved monotonicity properties compared to the case of indivisible goods. 

\begin{proposition}
\label{pro:APSdivisible}
    For allocation of a divisible homogeneous good, APS is both globally monotone and individually monotone.
\end{proposition}

\begin{proof}
    As the proportional allocation is acceptable by APS and globally monotone, this selection rule implies that APS is globally monotone. If valuation functions are strictly increasing, then the proportional allocation is the only acceptable allocation, which implies also individual monotonicity.

    We now consider individual monotonicity in the case that valuations might not be strictly increasing.  Suppose that $b_i >_i b'_i$. There is an allocation acceptable under $b'$ (the proportional allocation) in which $i$ gets only $v_i(b'_i)$, whereas in every allocation acceptable under $b$ agent $i$ gets at least $v_i(b_i)$. Hence lower monotonicity holds. Upper monotonicity was proved in Proposition~\ref{pro:upperMonotone}. 
    By Definition~\ref{def:monotone}, the combination of upper and lower monotonicity implies individual monotonicy.
\end{proof}

The proof presented for Proposition~\ref{pro:APSdivisible} is specific to the APS. However, the proposition itself holds for all proper shares (and hence also for Prop). The more general case is handled in Propositions~\ref{pro:lowerMonotone} and~\ref{pro:globalMonotone}.

\begin{proposition}
\label{pro:lowerMonotone}
     For allocation of a divisible homogeneous good, every fairness notion based on a proper share is lower monotone. Consequently, it is individually monotone.
\end{proposition}

\begin{proof}
Let $s$ be a proper share function (with arguments $v_i$ and $b_i$). Consider any two allocation instances that differ only in the entitlements, with $b >_i b'$. 
Let $t_i  = \min_{\{x \; \mid \; v_i(x) \ge  s(v_i,b_i)\}}[v_i(x)]$ denote the minimum of all values that agent $i$ might get in an allocation that is acceptable under $b_i$. In particular, if $v_i$ is continuous then $t_i = s(v_i,b_i)$. (If $v_i$ is not continuous then we might need to replace minimum by infimum. This adds some technical complications to the presentation of the proof, which are omitted here for lack of interest.)
If there is an allocation acceptable under $b'$ in which agent $i$ gets a value larger than $t_i$, there is also such an allocation that gives agent $i$ a value of exactly $t_i$ (by transferring a portion of the divisible good from agent $i$ to other agents). No allocation acceptable under $b$ can give agent $i$ a value lower than $t_i$. Hence, the fairness notion based on $s$ is lower monotone. Upper monotonicity was proved in Proposition~\ref{pro:upperMonotone}. The combination of the two implies individual monotonicity.   
\end{proof}

\begin{proposition}
\label{pro:globalMonotone}
     For allocation of a divisible homogeneous good, every fairness notion based on a proper share is globally monotone. 
\end{proposition}

\begin{proof}
Let $s$ be a proper share function (with arguments $v_i$ and $b_i$). We present a selection rule that is consistent with $s$ and is globally monotone. Consider an arbitrary allocation instance. For each agent $i$, let $x_i$ denote the smallest fraction of the good that gives her value of at least $s(v_i,b_i)$ (her share value). If $v_i$ is not continuous, then we take $x_i$ to be the infimum fraction rather than minimum. 
The selection rule tentatively gives each agent $i$ a fraction of the good equal to $f_i = x_i + \frac{1}{n}(1 - \sum_{i=1}^n x_i)$. If there is an acceptable allocation, then this allocation is acceptable.  If there is no acceptable allocation, then some of the $f_i$ might be negative, and hence the tentative allocation is not a legal allocation. To fix this, consider the operation {\em fix} which receives as an argument an agent $j$. If $f_j \ge 0$, then $fix(j)$ does nothing. If $f_j < 0$, then the negative $f_j$ is distributed equally among those agents who currently receive a positive fraction, updating the fraction that they receive (which might now become negative), whereas the new $f_j$ is set to~0. We repeat $fix(j)$ in a Round Robin fashion over all agents $j$ (this particular order is given only for concreteness -- end result will be the same regardless of the order), until no received fraction remains negative. As after each round, there is (at least) one more agent whose received fraction becomes~0, the fixing process must end. 

We have seen that the selection rule is consistent with the proper share $s$. We now show that it is monotone. Consider two instances that differ only in there entitlement, with $b >_i b'$. We need to show that under $b$ agent $i$ gets at least the same fraction of the good as under $b'$. This holds in the tentative allocation, because agent $i$ is the only one with $t_i \ge t'_i$ (this holds because the share is proper, and need not hold for a general share).  Therefore, for every other agent $j$, $t_i - t_j \le t'_i - t'_j$. This implies that at the tentative allocation we have two properties that hold:

\begin{itemize}
    \item $f_i - f_j \ge f'_i - f'_j$ for every agent $j$.
    \item $f_i \ge f'_i$.
\end{itemize}

Thereafter, one can show that after each fixing step, either it becomes that  $f_i=0$ (in which case it is implied that also $f'_i = 0$), or the two properties continue to hold. 
\end{proof}

CE is clearly both globally and individually monotone if valuation functions are strictly monotone (the proportional allocation is then the unique acceptable allocation). When valuations are not strictly monotone, we are able to prove only a weaker statement.

\begin{proposition}
\label{pro:CEdivisible}
    For allocation of a divisible homogeneous good, CE is globally monotone and lower monotone.
\end{proposition}

\begin{proof}
    Global monotonicity and lower monotonicity follow by exactly the same proof used for demonstrating these properties for the APS (Proposition~\ref{pro:APSdivisible}).
\end{proof}

To prove that CE is individually monotone, we need to prove that it is upper monotone. Though it can easily be seen to hold when valuation functions are strictly monotone, we encountered difficulties proving this when valuation functions are only weakly monotone.


For the remaining fairness notions, we present examples showing that they are not monotone.  In our examples we shall often chose valuation functions that are step functions, and consequently, they are only weakly increasing and non-continuous. This is done only for simplicity of the presentation. In all cases, the step functions can be replaced by strictly monotone continuous valuations without affecting the results.

WMMS is a general share but not a proper share, and so Propositions~\ref{pro:lowerMonotone} and~\ref{pro:globalMonotone} do not apply to it. In fact, unlike proper shares, it is neither individually monotone nor globally monotone. Moreover, 
it suffers from the incentive paradox.

\begin{proposition}
\label{pro:WMMSdivisible}
    For allocation of a divisible homogeneous good, WMMS is lower monotone, but suffers from the incentive paradox.
\end{proposition}

\begin{proof}
For WMMS, suppose that $b >_i b'$. Let $t$ and $t'$ denote the value of the WMMS of agent $i$ under $b$ and $b'$ respectively. Then $t \ge t'$. If there is an allocation acceptable under $b'$ in which agent $i$ gets a value larger than $t$, there is also such an allocation that gives agent $i$ a value of $t$ (by transferring a portion of the divisible good from agent $i$ to other agents). Hence WMMS is lower monotone.

We now present an example in which WMMS suffers from the incentive paradox. There are three agents and their valuations are presented in Table~\ref{tab:WMMSIncentive1}.

\begin{table}[hbt!]
        \centering
        \begin{tabular}{c|ccccc}
               & $0 < x < \frac{1}{4}$ & $\frac{1}{4} \le x < \frac{3}{8}$ &  $\frac{3}{8} \le x < \frac{1}{2}$ & $\frac{1}{2} \le x < \frac{5}{8}$ & $\frac{5}{8} \le x \le  1$\\
               \hline
            $v_1$ & 0 & 1 & 1 & 1 & 2   \\
                           \hline
            $v_2$ & $x$ & $x$ & $x$ & $x$ &  $x$   \\
                           \hline
            $v_3$ & 1 & 1 & 2 & 6 & 6  \\
        \end{tabular}
        \caption{Incentive paradox for WMMS, homogeneous divisible good.}
        \label{tab:WMMSIncentive1}
        \end{table}

 With entitlements $b' = (\frac{1}{2}, \frac{1}{4}, \frac{1}{4})$ the WMMS of agent~1 is~1 (in a WMMS partition of positive value, every agent must get more than a quarter of the good), the WMMS of agent~2 is $\frac{1}{4}$ (as her valuation is additive), and the WMMS of agent~3 is~1 (as it is impossible for all three agents to get at least $\frac{3}{8}$ fraction of the good). There are WMMS allocations in which agent~1 gets a value of~2, e.g., when the fractions allocated are $(\frac{5}{8}, \frac{1}{4}, \frac{1}{8})$. With entitlements $b = (\frac{5}{8}, \frac{1}{8}, \frac{1}{4})$ the WMMS of agent~1 is still~1, the WMMS of agent~2 is $\frac{1}{8}$, and the WMMS of agent~3 increases to~2 (due to the WMMS partition $(\frac{1}{2}, \frac{1}{8}, \frac{3}{8})$). In every allocation that gives agents~2 and~3 their WMMS, agent~1 can get at most half of the good, for a value of~1. Hence despite the fact that $b >_1 b'$, agent~1 is now limited to the smaller of the two values that she could get under $b'$. 
\end{proof}

The following proposition shows that despite the fact that the WMMS does not suffer from the inversion paradox (as it is lower monotone), it is not globally monotone.

\begin{proposition}
\label{pro:WMMSglobal}
    For allocation of a divisible homogeneous good, WMMS is not globally monotone.
\end{proposition}

\begin{proof}
We consider an instance with for agents. For this instance, we shall consider five vectors of entitlements.

\begin{itemize}
   \item $b = (0.1, 0.1, 0.4, 0.4)$. 
   \item $b^1 = (0.2, 0.1, 0.3, 0.4)$.
   \item $b^2 = (0.1, 0.2, 0.3, 0.4)$.
   \item $b^3 = (0.01, 0.1, 0.49, 0.4)$.
   \item $b^4 = (0.01, 0.1, 0.4, 0.49)$.
\end{itemize}

Observe that for every agent $i$ it holds that $b^i >_i b$. We shall design valuation functions for the agents that have the following properties.

\begin{itemize}
    \item For each $i \in \{1,2,3,4\}$, the vector $b^i$ of entitlements has a unique allocation acceptable acceptable under WMMS. In this allocation, the fraction of the good received by agent $i$ is strictly smaller than $b_i$, and of value lower than $v_i(b_i)$.
    \item For the vector $b$ of entitlements, there is an allocation acceptable under WMMS.
\end{itemize}

The above implies that there is no globally monotone selection rule consistent with WMMS. Such a selection rule must select an acceptable allocation for $b$, in this allocation some agent $i$ must receive at least a $b_i$ fraction of the good, but then monotonicity is violated when comparing to the allocation selected for $b^i$.

We proceed to describe the valuation functions. 
The valuations of agents~1 and~2 are identical, except for one of the steps of the step functions. Likewise,  The valuations of agents~3 and~4 are identical, except for one of the steps of the step functions. In our description of the valuation functions, we present the thresholds at which a new step begins, and the value of the function at that step (up to the next threshold). It is implicit in our description that for the step that starts at~0 the value of all functions are 0, and that the last step ends at~1. The value of the parameter $\delta$ can be set arbitrarily, provided that it is smaller than  $0.01$.

Table~\ref{tab:low} describes the valuations for agents~1 and~2.

\begin{table}[hbt!]
        \centering
        \begin{tabular}{c|ccccccc}
               & $\delta$ & $0.1 - 2\delta$ & $0.1 - \delta$ &  $0.1$ & $0.1+5\delta$ & $0.4+2\delta$ & $0.4+5\delta$\\
               \hline
            $v_1$ & 0 & 1 & 1.5 & 1.6 & 2  & 4 & 12  \\
                           \hline
            $v_2$ & $0.2$ & 1 & 1.5 & 1.6 & 2  & 4 & 12 \\
                           \hline
        \end{tabular}
        \caption{Valuations for low entitlement agents. At the noted threshold, the value increases to the value in the table.}
        \label{tab:low}
        \end{table}

Table~\ref{tab:high} describes the valuations for agents~3 and~4.

\begin{table}[hbt!]
        \centering
        \begin{tabular}{c|cccccccc}
               & $0.1 - 2\delta$ & $0.1 + 5\delta$ &  $0.1+10\delta$ & $0.4-20\delta$ & $0.4-2\delta$ & $0.4-\delta$ & $0.4$ & $0.4+5\delta$\\
               \hline
            $v_3$ & 1 & 2 & 2.5 & 2.5 & 4  & 4.9 & 4.95 & 5 \\
                           \hline
            $v_4$ & 1 & 2 & 2.5 & 3.9 & 4  & 4.9 & 4.95 & 5 \\
                           \hline
        \end{tabular}
        \caption{Valuations for high entitlement agents. At the noted threshold, the value increases to the value in the table.}
        \label{tab:high}
        \end{table}

To aide the reader in verifying the properties that were claimed to hold for these valuation functions we present tables~\ref{tab:agent1} to~\ref{tab:agent4}.  Each table describes an optimal WMMS partition for the respective agent for each of the entitlement vectors (this partition might not be unique), the ratio of value per entitlement implied by the partition, the value of the WMMS share for the agent, and the minimum fraction of the good that gives the agent this value. 


\begin{table}[hbt!]
        \centering
        \begin{tabular}{|c|cccc|c|c|c|}
        \hline 
               & \multicolumn{4}{|c|}{partition} & ratio & value &  fraction \\
               \hline
            $b$ & $0.1-2\delta$ & $0.1-2\delta$ & $0.4+2\delta$ & $0.4+2\delta$ & 10  & 1 & $0.1 - 2\delta$ \\
                           \hline
            $b^1$ & $0.1+5\delta$ & $0.1+5\delta$ & $0.1+5\delta$ & $0.4+5\delta$ & $\frac{2}{0.3}$  & $\frac{4}{3}$  & $0.1-\delta$ \\
            \hline
            $b^2$ & $0.1+5\delta$ & $0.1+5\delta$ & $0.1+5\delta$ & $0.4+5\delta$ & $\frac{2}{0.3}$  & $\frac{2}{3}$  & $0.1-2\delta$ \\
                           \hline
            $b^3$ & $0.1-2\delta$ & $0.1-2\delta$ & $0.4+2\delta$ & $0.4+2\delta$ & $1$  & $0.01$  & $0.1-2\delta$ \\
            \hline
            $b^4$ & $0.1-2\delta$ & $0.1-2\delta$ & $0.4+2\delta$ & $0.4+2\delta$ & $1$  & $0.01$  & $0.1-2\delta$ \\
                           \hline               
        \end{tabular}
        \caption{WMMS for agent~1, for each entitlement vector.}
        \label{tab:agent1}
        \end{table}

\begin{table}[hbt!]
        \centering
        \begin{tabular}{|c|cccc|c|c|c|}
        \hline 
               & \multicolumn{4}{|c|}{partition} & ratio & value &  fraction \\
               \hline
            $b$ & $0.1-2\delta$ & $0.1-2\delta$ & $0.4+2\delta$ & $0.4+2\delta$ & 10  & 1 & $0.1 - 2\delta$ \\
                           \hline
            $b^1$ & $0.1+5\delta$ & $0.1+5\delta$ & $0.1+5\delta$ & $0.4+5\delta$ & $\frac{2}{0.3}$  & $\frac{2}{3}$  & $0.1-2\delta$ \\
            \hline
            $b^2$ & $0.1+5\delta$ & $0.1+5\delta$ & $0.1+5\delta$ & $0.4+5\delta$ & $\frac{2}{0.3}$  & $\frac{4}{3}$  & $0.1-\delta$ \\
                           \hline
            $b^3$ & $\delta$ & $0.1+5\delta$ & $0.4+5\delta$ & $0.4+5\delta$ & $20$  & $2$  & $0.1+5\delta$ \\
            \hline
            $b^4$ & $\delta$ & $0.1+5\delta$ & $0.4+5\delta$ & $0.4+5\delta$ & $20$  & $2$  & $0.1+5\delta$ \\
                           \hline               
        \end{tabular}
        \caption{WMMS for agent~2, for each entitlement vector.}
        \label{tab:agent2}
        \end{table}

\begin{table}[hbt!]
        \centering
        \begin{tabular}{|c|cccc|c|c|c|}
        \hline 
               & \multicolumn{4}{|c|}{partition} & ratio & value &  fraction \\
               \hline
            $b$ & $0.1$ & $0.1$ & $0.4$ & $0.4$ & 10  & 4 & $0.4 - 2\delta$ \\
                           \hline
            $b^1$ & $0.1+5\delta$ & $0.1-2\delta$ & $0.4-2\delta$ & $0.4-\delta$ & 10  & 3  & $0.4-2\delta$ \\
            \hline
            $b^2$ & $0.1-2\delta$ & $0.1+5\delta$ & $0.4-2\delta$ & $0.4-\delta$ & 10  & 3  & $0.4-2\delta$ \\
                           \hline
            $b^3$ & $0.1-2\delta$ & $0.1-2\delta$ & $0.4+5\delta$ & $0.4-\delta$ & $10$  & 4.9  & $0.4-\delta$ \\
            \hline
            $b^4$ & $0.1-2\delta$ & $0.1-2\delta$ & $0.4-\delta$ & $0.4+5\delta$ & $10$  & 4  & $0.4-2\delta$ \\
                           \hline               
        \end{tabular}
        \caption{WMMS for agent~3, for each entitlement vector.}
        \label{tab:agent3}
        \end{table}

        \begin{table}[hbt!]
        \centering
        \begin{tabular}{|c|cccc|c|c|c|}
        \hline 
               & \multicolumn{4}{|c|}{partition} & ratio & value &  fraction \\
               \hline
            $b$ & $0.1$ & $0.1$ & $0.4$ & $0.4$ & 10  & 4 & $0.4 - 2\delta$ \\
                           \hline
            $b^1$ & $0.1+10\delta$ & $0.1+5\delta$ & $0.4-20\delta$ & $0.4+5\delta$ & 12.5  & 5  & $0.4+5\delta$ \\
            \hline
            $b^2$ & $0.1+5\delta$ & $0.1+10\delta$ & $0.4-20\delta$ & $0.4+5\delta$ & 12.5  & 5  & $0.4+5\delta$ \\
                           \hline
            $b^3$ & $0.1-2\delta$ & $0.1-2\delta$ & $0.4+5\delta$ & $0.4-\delta$ & $10$  & 4  & $0.4-2\delta$ \\
            \hline
            $b^4$ & $0.1-2\delta$ & $0.1-2\delta$ & $0.4-\delta$ & $0.4+5\delta$ & $10$  & 4.9  & $0.4-\delta$ \\
                           \hline               
        \end{tabular}
        \caption{WMMS for agent~4, for each entitlement vector.}
        \label{tab:agent4}
        \end{table}
\end{proof}

The remaining two fairness notions that we consider suffer from the inversion paradox. 

\begin{proposition}
    For allocation of a divisible homogeneous good, WEF suffers from the inversion paradox.
\end{proposition}

\begin{proof}
    Consider an example with three agents and valuations as in Table~\ref{tab:WEFIversion1}. 

    \begin{table}[hbt!]
        \centering
        \begin{tabular}{c|ccccccc}
               & $x < .2$ & $x = .2$ &  $.2 < x \le .25$ & $.25 < x < .5$ & $.5 \le x < .55$ & $x \ge .55$\\
               \hline
            $v_1$ & 0 & 0 & 4 & 5 & 5 & 5 \\
                           \hline
            $v_2$ & 0 & 1 & 1 & 1 &  1  &  5\\
                           \hline
            $v_3$ & 0 & 2 & 2 & 2 & 5  & 10\\
        \end{tabular}
        \caption{Inversion paradox for WEF, homogeneous divisible good.}
        \label{tab:WEFIversion1}
        \end{table}

With entitlements $b'=(0.3, 0.2, 0.5)$, the WEF allocation $(0.3,0.2,0.5)$ gives agent~1 a value of~5. No WEF allocation gives agent~1 a lower value, because in every WEF allocation the fraction given to agent~2 is $0.2$ (at higher value, agent~1 will envy agent~2 in a weighted sense), the fraction given to agent~3 is strictly less than $0.55$ (so that agent~2 does not envy agent~3), and so the fraction given to agent~1 is strictly more than $0.25$. With entitlements $b=(0.4,0.1,0.5)$ (satisfying $b >_1 b'$) the WEF allocation $(0.25,0.2,0.55)$ gives agent~1 a value of only~4. No WEF allocation gives agent~1 a higher value, because in every WEF allocation the fraction given to agent~2 is $0.2$, the fraction given to agent~3 is at least $0.55$ (so that agent~3 does not envy agent~2 in a weighted sense), and so the fraction given to agent~1 is at most $0.25$.
\end{proof}

\begin{proposition}
\label{pro:MWNSWdivisible}
    For allocation of a divisible homogeneous good, MWNSW suffers from the inversion paradox. 
\end{proposition}

\begin{proof}
We show that when valuations are not concave, MWNSW sometimes suffers from the inversion paradox. Consider an instance with three agents. Let $k$ be a sufficiently large constant. The value of a fraction $x$ of the good is shown in Table~\ref{tab:MWNSWInversion1}. 

\begin{table}[hbt!]
        \centering
        \begin{tabular}{c|ccccc}
               & $0 < x < \frac{1}{6}$ & $\frac{1}{6} \le x < \frac{2}{6}$ &  $\frac{2}{6} \le x \le \frac{4}{6}$ & $\frac{4}{6} < x <  1$\\
               \hline
            $v_1$ & $\frac{1}{k}$ & 1 & 2 & 3   \\
                           \hline
            $v_2$ & $\frac{1}{k}$ & 1 & 2 & 3    \\
                           \hline
            $v_3$ & $\frac{1}{k}$ & 1 & 2 &   $k$  \\
        \end{tabular}
        \caption{Inversion paradox for MWNSW, homogeneous divisible good.}
        \label{tab:MWNSWInversion1}
        \end{table}

With equal entitlements the allocation that maximizes NSW gives every agent $\frac{1}{3}$, for a NSW value of~2. Increasing the entitlement of agent~1 to nearly $\frac{2}{3}$ and making the entitlement of agent~2 negligible, NSW is maximized by giving agent~3 strictly more than $\frac{2}{3}$, giving agent~2 a negligible fraction, and giving agent~1 a fraction of value~1. Agent~1 suffers from the inversion paradox.    
\end{proof}

We summarize our results concerning monotonicity for a divisible homogeneous good in Table~\ref{tab:monotonicityDivisible}.

\begin{table}[hbt!]
        \centering
\begin{tabular}{ | m{2cm} || m{2cm}| m{2cm} | m{2cm} | m{2cm} |} 
 \hline
 \multicolumn{5}{|c|}{Homogeneous Divisible Good} \\
 \hline
  Fairness \newline notion & global \newline monotonicity & individual \newline monotonicity & incentive \newline paradox & inversion \newline paradox\\ 
 \hline
 \hline
 APS & yes & yes & immune & immune\\
 Prop & yes & yes & immune & immune\\ 
 CE & yes & ? & ? & immune\\ 
  \hline
 WMMS & no & no$^*$ & suffers & immune \\ 
  \hline
 WEF & no & no & suffers & suffers \\ 
 MWNSW & no & no & suffers & suffers \\ 

 \hline
\end{tabular}
        \caption{The examples showing the paradoxes are with valuation functions that are not concave. CE is lower monotone, but we do not know if it is upper monotone when valuations are only weakly monotone. $(^*)$ WMMS is lower monotone but not upper monotone.}
        \label{tab:monotonicityDivisible}
        \end{table}

In the interesting special case in which all valuations are concave, MWNSW is both globally and individually monotone and does not suffer from any of the paradoxes. See Proposition~\ref{pro:MWNSWconcave} in Section~\ref{sec:MWNSWconcave}.

\section{Attitude towards welfare}
\label{sec:welfare}

In this section we discuss the attitude of the fairness notions towards social welfare. We shall address two aspects -- that of Pareto optimality, and that of Weighted Nash Social Welfare (WNSW).

\subsection{Pareto optimality}
\label{sec:pareto}

Here we consider three classes of fairness notions as defined in Section~\ref{sec:attituteDef}: Pareto, pro-Pareto and non-Pareto. We shall consider two settings. One is that of indivisible goods, and the other is that of a divisible homogeneous good and valuation function that are weakly but not strictly increasing. The setting of a divisible homogeneous good and valuation functions that are strictly increasing need not be considered, because in this case every allocation is Pareto optimal (and hence all fairness notions are classified as Pareto). 

It is easy to see and well known that share based fairness notions (Prop, APS, WMMS, in our case) are pro-Pareto. 

MWNSW belongs to the class Pareto. 
(A minor technical issue is non-continuous valuation functions when allocating a divisible homogeneous good, in which case one might arfue that a MWNSW allocation need not exist. However, as we consider only acceptable allocations, the non-existence of such allocations frees us from the need that they be Pareto optimal.)

The following proposition establishes that WEF is non-Pareto. The proposition was known prior to this work, and we present a proof only for completeness. 

\begin{proposition}
    There are allocation instances with equal entitlements that have envy free allocation, but for which no envy free allocation is Pareto optimal.
\end{proposition}

\begin{proof}
    Consider a divisible Homogeneous good, with two agents with the following valuation functions.

\begin{table}[hbt!]
        \centering
        \begin{tabular}{c|cccc}
               & $0 < x < \frac{1}{4}$ & $\frac{1}{4} \le x < \frac{1}{2}$ &   $\frac{1}{2} \le x <  1$\\
               \hline
            $v_1$ & $x$ & $x$ & $x$    \\
                           \hline
            $v_2$ & $2x$ & $\frac{1}{2}$ & $x$    \\
        \end{tabular}
        \caption{Non-Pareto WEF, homogeneous divisible good.}
        \label{tab:WEFPareto}
        \end{table}

With equal entitlements the only WEF allocation is $(\frac{1}{2},\frac{1}{2})$, but it is Pareto dominated by the allocation $(\frac{3}{4},\frac{1}{4})$.

For the case of indivisible goods, consider five items and three agents with equal entitlements and the following valuation functions.

 \begin{table}[hbt!]
        \centering
        \begin{tabular}{c|ccccc}
               & $e_1$ & $e_2$ & $e_3$ & $e_4$ & $e_5$ \\
               \hline
            $v_1$ & 5 & 4 & 3 & 2 & 1 \\
                           \hline
            $v_2$ & 4 & 3 & 3 & 2 & 2 \\
                           \hline
            $v_3$ & 4 & 2 & 2 & 3 & 3 \\
        \end{tabular}
        \caption{Non-Pareto WEF, indivisible goods.}
        \label{tab:WEFPareto1}
    \end{table}

This instance has two WEF allocations $(\{e_1\}, \{e_2, e_5\}, \{e_3, e_4\})$ and $(\{e_1\}, \{e_3, e_4\}, \{e_2, e_5\})$. Both of them are Pareto dominated by the allocation $(\{e_1\}, \{e_2, e_3\}, \{e_4, e_5\})$.
\end{proof}

For CE, it is stated in~\cite{BNT19} and in \cite{SegalHalevi20} that every CE allocation is Pareto optimal. This indeed holds when all valuations are ``strict" (no two bundles have the same value), as shown for example in \cite{BNT21}. However, in general, CE allocations need not be Pareto optimal, a fact that was already noted in earlier works~\cite{MasColell}.

\begin{proposition}
    There are instances in which a competitive equilibrium exists, yet no Pareto optimal allocation has prices under which it is a competitive equilibrium. In other words, CE belongs to the class non-Pareto. 
\end{proposition}

\begin{proof}
For a divisible homogeneous good, the same example that shows that WEF is non-Pareto (Table~\ref{tab:WEFPareto}) also shows that CE is non-Pareto. (With equal entitlement, given any pricing for portions of the good, if agent~1 can afford more than half the good, so can agent~2.) 

For indivisible goods, consider an instance with three items and three agents, with the following additive valuation functions.

\begin{table}[hbt!]
        \centering
        \begin{tabular}{c|ccc}
               & $e_1$ & $e_2$ & $e_3$  \\
               \hline
            $v_1$ & 1 & 1 & 1  \\
                           \hline
            $v_2$ & 2 & 1 & 1  \\
                           \hline
            $v_3$ & 2 & 1 & 1  \\
        \end{tabular}
        \caption{Non-Pareto CE, indivisible goods.}
        \label{tab:CEPareto1}
    \end{table}

With entitlements $b = (0.4, 0.3, 0.3)$, the only allocations acceptable by CE are $(\{e_1\}, \{e_2\}, \{e_3\})$ and $(\{e_1\}, \{e_3\}, \{e_2\})$, supported by the price vector $p=(0.4,0.3,0.3)$. Neither one of them is Pareto optimal (swapping $e_1$ and $e_2$ gives allocations that Pareto dominate them). 
\end{proof}

The results of this section are summarized in Table~\ref{tab:pareto}.

\begin{table}[hbt!]
        \centering
\begin{tabular}{ | m{3cm} || m{3cm} |} 
 \hline
Fairness notion & Pareto class \\ 
\hline
\hline
 MWNSW & Pareto  \\ 
  \hline
  Prop & pro-Pareto \\ 
 APS & pro-Pareto \\
 WMMS & pro-Pareto  \\ 
 \hline
 CE & non-Pareto \\  
 WEF & non-Pareto  \\ 
 \hline
\end{tabular}
        \caption{The examples showing non-Pareto are with additive valuations and indivisible goods, and with weakly monotone valuations and a divisible homogeneous good. 
        }
        \label{tab:pareto}
        \end{table}

\subsection{Attitute towards WNSW}
\label{sec:WNSW}

In this section we consider a divisible homogeneous good. We classify fairness notions according to the deviations that they impose from the proportional allocation. The four classes, neutral, risky-WNSW, pro-WNSW and non-WNSW are defined in Section~\ref{sec:attituteDef}.

As the good is divisible and homogeneous, there is only one possible linear valuation function (up to scaling).  When all agents have this linear valuation, all fairness notions considered in this manuscript coincide. 

\begin{proposition}
\label{pro:linear}
    If the $n$ agents have arbitrary entitlements and linear valuation functions, then for all the following fairness notions, there is a unique acceptable allocation, which is the proportional allocation (that gives each agent $i$ a $b_i$-fraction of the divisible good).
    \begin{itemize}
        \item Share based notions: Prop, APS, WMMS.
        \item Envy based notions: WEF.
        \item Optimization based notions: MWNSW.
        \item Equilibrium based notions: CE.
    \end{itemize}
\end{proposition}

\begin{proof}
    All parts of the proposition follow quite easily from the respective definitions. Here we present the proof only for the case of MWNSW. By scaling (which is allowed, as the MWNSW allocation remains unchanged when valuation functions are scaled), we may assume that all agents have the same valuation function $f(x) = x$. We first consider the case of $n=2$. Let the entitlements be $b_1 = b$ and $b_2 = 1-b$, and consider an allocation $(x,1-x)$. The weighted Nash Social Welfare is $x^b (1-x)^{1-b}$. Taking logarithms we have $b\log x + (1-b)\log (1-x)$. To see for which value of $x$ this expression is maximized, we take derivatives with respect to $x$ and get $\frac{b}{x} - \frac{1-b}{1-x}$. The derivative is~0 when $x=b$, namely, the proportional allocation. For $n > 2$, an argument similar to the above shows that for every pair of agents, the ratio between the fractions that they receive from the good must be equal to the ratio between their entitlements. This can hold only for the proportional allocation.
\end{proof}

We now consider valuation functions that need not be linear. We observe that for the APS and for CE, the proportional allocation is acceptable, and in every acceptable allocation, every agent gets value at least as large as in the proportional allocation. Hence these notions are neutral according to our classification. MWNSW is clearly pro-WNSW.  The classification for Prop is more interesting. If valuation functions are concave, then the proportional allocation is acceptable, and if at least one of the valuation functions is strictly concave,  then other allocations are acceptable as well. It is not difficult to show examples in which some of these other acceptable allocations  have lower WNSW than that of the proportional allocation, and some have higher WNSW. If valuation functions are convex and at least one of them is strictly convex, then there are no acceptable allocations at all. The more general case that allows for both concavity and convexity is handled by the following proposition.

\begin{proposition}
For allocation of a divisible homogeneous good, there are instances with two agents and identical valuations in which there is a unique allocation that gives each agent her proportional share, and this allocation has lower WNSW than the proportional allocation.
\end{proposition}

\begin{proof}
There are two agents with entitlements $b = (\frac{1}{4}, \frac{3}{4})$ and the same piece-wise linear continuous valuation function $v$.  (Observe that $v$ is convex in the range $[0, \frac{3}{4}]$ and then becomes flat, so overall it is neither concave not convex.)

\begin{table}[hbt!]
        \centering
        \begin{tabular}{c|ccccc}
               & $0 < x \le \frac{1}{3}$ & $\frac{1}{3} \le x \le \frac{2}{3}$ &  $\frac{2}{3} \le x \le \frac{3}{4}$ & $\frac{3}{4} \le x \le  1$\\
               \hline
            $v$ & $\frac{3}{4}x$ & $\frac{3}{2}x - \frac{1}{4}$ & $3x - \frac{5}{4}$ & 1   \\
        \end{tabular}
        \caption{Valuation for which Prop lowers WNSW.}
        \label{tab:ProportionalWNSW}
        \end{table}

The WNSW of the proportional allocation $(\frac{1}{4}, \frac{3}{4})$ is $(\frac{3}{16})^{\frac{1}{4}} \cdot 1^{\frac{3}{4}} = \frac{1}{4}\cdot (48)^{\frac{1}{4}}$. The unique allocation that gives each agent at least her proportional share is $(\frac{1}{3}, \frac{2}{3})$. It has lower WNSW, $(\frac{1}{4})^{\frac{1}{4}} \cdot (\frac{3}{4})^{\frac{3}{4}}= \frac{1}{4}\cdot (27)^{\frac{1}{4}}$.
\end{proof}

For the remaining two fairness notions (WMMS and WEF),
results will not be presented in full generality, but rather only in the special case of two agents with entitlements $b_1 = \frac{1}{3}$ and $b_2 = \frac{2}{3}$, and with valuations that are either $v(x) = x$ (linear), $v(x) = \sqrt{x}$ (a representative example of a strictly concave function), $v(x) = x^2$ (a representative example of a strictly convex function). This special case suffices in order to illustrate the main points that we wish to make, and offering a more comprehensive classification does not appear to serve a useful purpose (in the context of the current paper).

\begin{proposition}
    \label{pro:antiNSW}
    Suppose that there are two agents with entitlements $b_1 = \frac{1}{3}$ and $b_2 = \frac{2}{3}$. If both agents have the same valuation function and it is either $v(x) = \sqrt{x}$ or $v(x) = x^2$, then for the following fairness notions, there is a unique acceptable allocation, and it has strictly smaller weighted Nash Social Welfare (WNSW) than the proportional allocation.
\begin{itemize}
        \item Share based notions: WMMS.
        \item Envy based notions: WEF.
    \end{itemize}
    \end{proposition}

\begin{proof}
If both agents have the same valuation function, then with entitlements $b_1 = \frac{1}{3}$ and $b_2 = \frac{2}{3}$, in an WEF allocation agent~2 should get twice as high value as agent~1. Likewise, the partition that attains the maximum in Definition~\ref{def:WMMS} of the WMMS is one in which  agent~2 gets twice as high value as agent~1. Hence the WMMS and WEF allocations are identical. 

    For $v(x) = \sqrt{x}$ the WEF/WMMS allocation is $(\frac{1}{5},\frac{4}{5})$, which satisfies $2\sqrt{\frac{1}{5}} = \sqrt{\frac{4}{5}}$. The NSW is $\left(\frac{1}{5}\right)^{1/6} \cdot \left(\frac{4}{5}\right)^{2/6} < 0.71$. The NSW of the proportional allocation is  $\left(\frac{1}{3}\right)^{1/6} \cdot \left(\frac{2}{3}\right)^{2/6} > 0.72$.

    For $v(x) = x^2$ the WEF/WMMS allocation is $(\sqrt{2}-1,\; 2-\sqrt{2})$, which satisfies $2(\sqrt{2}-1)^2 = (2-\sqrt{2})^2$. The NSW is $\left(\sqrt{2}-1\right)^{2/3} \cdot \left(2-\sqrt{2}\right)^{4/3} < 0.273$. The NSW of the proportional allocation is  $\left(\frac{1}{3}\right)^{2/3} \cdot \left(\frac{2}{3}\right)^{4/3} > 0.279$.
\end{proof}

The classification with respect to attitude towards WNSW is presented in Table~\ref{tab:NashInversion}.

\begin{table}[hbt!]
        \centering
\begin{tabular}{ | m{2cm} || m{2cm}| m{2cm} | m{2cm} | m{2cm} |} 
 \hline
 \multicolumn{5}{|c|}{WNSW for Homogeneous Divisible Good} \\
 \hline
  Fairness \newline notion & additive \newline valuations & concave \newline valuations & convex \newline valuations & arbitrary \newline valuations\\ 
 \hline
 \hline
 APS & neutral & neutral & neutral & neutral\\

 CE & neutral & neutral & neutral & neutral\\ 
  \hline
 Prop & neutral & risky & not feasible & non \\ 
 
  \hline
  WMMS & neutral & non & non & non \\
 WEF & neutral & non & non & non \\
 \hline
 MWNSW & neutral & pro & pro & pro \\ 

 \hline
\end{tabular}
        \caption{{\em Neutral} means that the proportional allocation is acceptable, and if there are additional acceptable allocations, they  have at least as high WNSW. {\em Risky} means that the proportional allocation is acceptable, there typically are other acceptable allocations, and their WNSW might be either higher or lower. {\em Pro} means that WNSW is never smaller than that of the proportional allocation, and typically larger. {\em Non} means that on some instances, in every acceptable allocation the WNSW is smaller than that of the proportional allocation.}
        \label{tab:NashInversion}
        \end{table}

\section{Discussion}
\label{sec:discussion}

For every given allocation instance, a fairness notion partitions the set of allocations into two subsets (one of which might be empty) -- those that are acceptable under the fairness notions, and those that are not. A minimal interpretation of the fairness notion only postulates that if the set of acceptable allocations is not empty, then the selected allocation needs to be one of the acceptable allocations. In this paper, we investigate the implications of this interpretation, for six previously studied fairness notions (Prop, APS, WMMS, WEF, CE, MWNSW). All these fairness notions share some common features: they allow for agents with possibly unequal entitlements, the acceptable sets depend only on the entitlements of the agents and on their valuation functions, and they are scale invariant (the set of acceptable allocations does not change if a valuation function is scaled by a multiplicative positive constant). We considered two settings. One was that of a homogeneous divisible good, and the valuation functions could be arbitrary (though recall that we require valuations to be non-decreasing). The other was that of indivisible items, with special emphasis on additive valuations.

\subsection{Monotonicity}

The main property of interest in this paper is that of monotonicity. We studied two forms of monotonicity: global monotonicity and individual monotonicity (which combines upper monotonicity and lower monotonicity). We also introduced the inversion paradox which contradicts both these monotonicity properties, and the weaker incentive paradox that contradicts individual monotonicity.

A central finding of this work is that all fairness notions based on proper shares (these include Prop and APS) enjoy some level of monotonicity: they are upper monotone (and hence do not suffer from the incentive paradox), and in the case of a divisible homogenous good, they are both globally monotone and individually monotone. All other fairness notions that we considered (WMMS, which is a share but not a proper share, WEF, CE, MWNSW) do suffer from the incentive paradox. Moreover, even in the case of a divisible homogeneous good, WEF and MWNSW suffer from the incentive paradox, whereas WMMS suffers from the incentive paradox and is not globally monotone. 

Our results imply that none of our fairness notions are globally monotone, except possibly for the APS. 

\begin{question}
\label{que:global}
    Is the APS globally monotone? Is it globally monotone in the special case of additive valuations?
\end{question}

If the answer to the above question turns out to be negative, then one may ask whether there is any natural fairness notion (one not considered in the current paper) that is globally monotone. Here is an example of one such fairness notion. We say that an allocation $A = (A_1, \ldots, A_n)$ has {\em no downward envy} (NDE) if for every pair $(i,j)$ of agents, if $b_i > b_j$, then $v_i(A_i) \ge v_i(A_j)$. In other words, an agent does not envy an agent that has strictly lower entitlement. The following selection rule shows that the NDE fairness notion is globally monotone: allocate all of $\items$ to the agent of highest entitlement (breaking ties arbitrarily). Of course, the fact that this selection rule is globally monotone and consistent with the NDE fairness notion should not be interpreted as an endorsement to use this selection rule.

As to the incentive paradox, all our fairness notions suffer from it in the case of indivisible goods. However, for APS, our negative example involves a valuation function that is not additive. This leaves the hope that APS does not suffer from the inversion paradox when valuations are additive. 

\begin{question}
\label{que:APS}
    Does the APS suffer from the incentive paradox when valuations are additive?
\end{question}

\subsection{Size-based versus value-based fairness notions}
\label{sec:sizevalue}

Consider two agents $i$ and $j$ with different entitlements $b_i$ and $b_j$, where $b_i = r\cdot b_j$ for some $r > 1$. Our study for the case of a divisible homogeneous good clarifies that different fairness notions treat this ratio $r$ in different ways. We classify these different treatments into three classes.

\begin{itemize}
    \item {\em Size based.} Agent $i$ deserves a piece of the good whose size is $r$ times as large as the piece received by agent $j$. This leads to the proportional allocation. Under this interpretation, the entitlement of the agents are to certain fractions of the good, and each agent should receive the fraction that she is entitled to. For the purpose of fairness, the valuations of agents do not matter. Valuations may matter for the purpose of economic efficiency, but this is a different issue, not related to fairness. Valuations may (and do) matter also when the good is not homogeneous (including the case of indivisible items), because there the valuation functions can be used as an interpretation of what constitutes a $b_i$ fraction of the good (an interpretation that is not needed when the good is homogeneous). APS and CE belong to this class.
    \item {\em Value based.} Agent $i$ deserves a piece of the good whose value is $r$ times as large as the value received by agent $j$. As values are subjective, the set of acceptable allocations depends on the valuations of the agents, and is sometimes empty. We note that in value based fairness notions, it need not be the case that valuations are used in order to improve economic efficiency. For example, our classification in Table~\ref{tab:NashInversion} shows that these fairness notions often decrease weighted Nash Social Welfare.  Prop, WMMS and WEF belong to this class.
    \item {\em Optimization based.} Here there is no direct interpretation for the ratio $r$. Rather, $b_i$ and $b_j$ appear as parameters in some objective, and an allocation is acceptable if it optimizes the objective. MWNSW belongs to this class.
\end{itemize}

We end with a few comments on each of the fairness notions that we considered.

\subsection{The proportional share}

Being a proper share, Prop enjoys relatively good monotonicity properties. It is interesting to note that for the case of a divisible homogeneous good (and arbitrary valuations), the proportional allocation need not be acceptable under Prop. If one has in mind the proportional allocation, a share based notion that is compatible with it is the APS, not Prop. 

Though Prop is not a feasible share, it is feasible in expectation (giving agent $i$ the grand bundle $\items$ with probability $b_i$).

\subsection{The anyprice share}

Compared to other fairness notions considered in this paper, we find that the APS performs well with respect to the properties considered in this paper. A positive answer to any of the questions~\ref{que:global} or~\ref{que:APS} would further strengthen its standing.

\subsection{The weighted maximin share}

Not being a proper share, the WMMS does not enjoy the good monotonicity properties of proper shares. For indivisible goods, it suffers from the incentive paradox. Importantly, this holds even when agents have additive valuations, which is the class of valuations considered in the paper that introduced the WMMS~\cite{FGHLPSSY19}. For non-additive valuations, WMMS appears to perform quite erratically even in the case of a divisible homogeneous good: it is not always feasible, and when it is, it often has lower weighted Nash social welfare compared to the natural proportional allocation.

There is one aspect in which the WMMS has an advantage over the other fairness notions considered in this work (except for MWNSW), and that is the fact that it is feasible in the case of an external valuation function (when all valuation functions are identical). This (and other considerations) motivated the author to ask whether some natural proper share notion (recall that the WMMS is not a proper share) is feasible when there is an external valuation function (and entitlements need not be equal). For the share notions that the author considered (the pessimistic share, and another notion referred to here as $MMS^-$), the answer turns out to be negative. See Section~\ref{sec:external} for more details. 

\subsection{Weighted envy-freeness}
\label{sec:disWEF}

In general, it appears that WEF performs quite poorly with respect to the properties considered in this paper. It suffers from the inversion paradox not only in the case of indivisible goods, but even in the case of a homogeneous divisible good. 

One might argue that for envy-based fairness notions, our monotonicity properties (as defined) are not very relevant. This is because an agent may be willing to suffer a loss in value, provided that other agents suffer even more. Though our formal definitions for monotonicity leave this option open, our negative examples do not. Inspecting the proof of Proposition~\ref{pro:notMonotone}, one sees that when the entitlement of agent~1 increases, the value that she receives decreases, whereas the values that all other agents receive increases. 

Another observation that we make is that unlike the case of equal entitlements, in which every CE allocation is envy free, for unequal entitlements, it is not true that CE implies WEF. This is evident from the case of a divisible homogeneous good, in which CE dictates the proportional allocation, whereas  WEF is not always feasible, and even when it is, the associated allocations change dramatically based on valuations of the agents.

For the case of a divisible homogeneous good, we considered non-additive valuations. There was earlier work~\cite{CSS22, MSST22} that suggested that the notion of WEF might not be appropriate for non-additive valuations, and proposed alternative notions. The motivation in these earlier works appears to be issues of feasibily, or issues of feasibility of relaxed versions of WEF (such as WEF1), and not issues of monotonicity (that manifest themselves even when agents have additive valuations).

One may argue that for allocation of a divisible homogeneous good to agents with arbitrary entitlements, the proportional allocation should be regarded as being ``envy free".
One may ask whether there is an extension of envy freeness to the arbitrary entitlement case under which this allocation is acceptable (WEF fails in this respect). Keeping in mind the classification of Section~\ref{sec:sizevalue}, this seems to require an extension that is size-based rather than value-based. Here we present such an extension, that we refer to as {\em Calibrated Envy Freeness} (CEF). It postulates that first agent $i$ computes what she believes to be the share value of each agent $j$, based on the entitlement of agent $j$ and the valuation of agent $i$.  Then she computes what fraction of this share value each agent received, and finds the allocation to be CEF if no other agent received a larger fraction than she herself did. So as to make CEF size-based, we use a size-based share, the APS. (Other size based shares can also be considered for this purpose. In contrast, if we use the value based share Prop instead of APS in this framework, the definition becomes equivalent to that of WEF.)

\begin{definition}
\label{def:CEF}
    An allocation $(A_1, \ldots, A_n)$ to agents of arbitrary entitlements (with entitlements $(b_1, \ldots, b_n)$ and valuations $(v_1, \ldots, v_n)$) is {\em calibrated envy-free} (CEF) if for every pair of agents $i$ and $j$, $\frac{v_i(A_i)}{APS(v_i,b_i)} \ge \frac{v_i(A_j)}{APS(v_i,b_j)}$. (Here $APS(v,b)$ is the APS value of an agent that has valuation function $v$ and entitlement $b$.)
\end{definition}

\begin{proposition}
    When allocating a divisible homogeneous good, the proportional allocation is acceptable under CEF.
\end{proposition}

\begin{proof}
The proportional allocation is CEF, because from the point of view of each agent $i$, she received value exactly equal to her APS, and every other agent $j$ received exactly the APS corresponding to entitlement $b_j$. Hence $\frac{v_i(A_i)}{APS(v_i,b_i)} = \frac{v_i(A_j)}{APS(v_i,b_j)} = 1$ for all agents $i$ and $j$.  
\end{proof}

\subsection{Competitive equilibrium}
\label{sec:disCE}

In our classification we left the following questions open (see Table~\ref{tab:monotonicityDivisible}): 

\begin{question}
\label{que:CE}
  For allocation of a divisible homogeneous good, if agents may have weakly monotone valuations, is CE upper monotone? Might CE suffer from the incentive paradox? 
\end{question}

We shall soon see a variation on this question for which we do know the answer.  Recall the notion of Parsimonious CE (PCE) from Remark~\ref{rem:PCE}.

\begin{definition}
\label{def:PCE}
    Given an allocation instance with $n$ agents with valuations $(v_1, \ldots, v_n)$ and entitlements $(b_1, \ldots, b_n)$, an allocation $(A_1, \ldots, A_n)$ is a {\em parsimonious competitive equilibrium} (PCE) if there are non-negative prices $(p_1, \ldots, p_m)$ to the items such that for every agent $i$, the bundle $A_i$ is the highest value bundle among the affordable bundles (according to $v_i$), {\em and of lowest price among these affordable highest value bundles}.  For divisible items, we assume that an $\alpha$-fraction of an item costs an $\alpha$-fraction of the total cost of the item. 
\end{definition}

Unlike CE, every PCE allocation is Pareto optimal. We present the proof for completeness. Suppose for the sake of contradiction that allocation $B = (B_1, \ldots B_n)$ Pareto dominates a PCE allocation $A = (A_1, \ldots, A_n)$. Then for some agent $i$, $v_i(B_i) > v_i(A_i)$. The fact that $A$ is a PCE implies that $B_i$ is priced higher than $A_i$. Consequently, for some $j$, $B_j$ is priced lower than $B_i$. The fact that $A$ is a PCE implies that $v_j(A_j) > v_j(B_j)$, contradicting the assumption that $B$ Pareto dominates $A$. (For CE the above argument fails because we can only conclude that $v_j(A_j) \ge v_j(B_j)$.)

On the one hand, the fact that PCE implies Pareto optimality and CE does not may serve as motivation to prefer PCE over CE. 

On the other hand, there may be arguments for preferring CE over PCE. One of them relates to feasibility: CE allows for more allocations. Every PCE allocation is CE but there are CE allocations (in particular, those that are not Pareto optimal) that are not PCE. In fact, for the case of a divisible homogeneous good, CE is feasible, whereas there are instances for which there is no acceptable allocation under PCE (for example, the instance in Table~\ref{tab:WEFPareto}). Another relates to the nature of the equilibrium. For CE, one may assume that every agent that spends some of her budget spends all of her budget (as we can raise the price of her received bundle so that it matches her budget). For PCE this is not true, as demostrated by an instance with two identical indivisible items and two agents with entitlements $b = (0.6,0.4)$ (the higher entitlement agent cannot spend all her budget in a PCE). 

CE and not PCE was the fairness notion considered in~\cite{BNT21} and some other related work. It is not clear to the author whether this was done as a deliberate choice, weighing the advantages and disadvantages of each option, or whether this was done because the authors were not aware of the distinction between CE and PCE. 

Considering Question~\ref{que:CE} with PCE instead of CE, we observe that for allocation of a divisible homogeneous good, the only allocation that might be acceptable under PCE is the proportional allocation. Consequently, PCE is individually monotone, and in particular, is upper monotone and does not suffer from the incentive paradox. Hence Question~\ref{que:CE} is not open if CE is replaced by PCE, suggesting to us that that the question is only of marginal interest.

Though the use of PCE instead of CE does offer some advantages in the context of this paper (it guarantees Pareto optimality, it allows us to fill up the missing cells in Table~\ref{tab:monotonicityDivisible}), it does not remedy the situation with respect to the inversion paradox. The example in Table~\ref{tab:CEInversion} that shows the inversion paradox for CE is applicable without change to PCE as well.

\subsection{Maximum weighted Nash social welfare}

According to our monotonicity properties, the performance of MWNSW is very poor. It suffers from the incentive paradox both in the case of a divisible homogeneous good, and in the case of indivisible goods (and additive valuations). We remind the reader that MWNSW does satisfy a weaker monotonicity property, {\em weight monotonicity} (see Section~\ref{sec:related}).

\subsection{Other fairness notions}

In this work, we studied the monotonicity properties of six fairness notions (Prop, APS, WMMS, WEF, MWNSW, CE). There are other fairness notions (and additional ones may be introduced in the future), and so our study is not exhaustive. Here we only wish to comment on one aspect of possible future research, and this concerns what we refer to as {\em relaxed fairness notions}.

We say that a fairness notion $R$ is a relaxation of fairness notion $F$ if every allocation that is acceptable under $F$ is also acceptable under $R$.

Depending on the nature of $R$ and $F$, there may be different interpretations of what the relaxation relation should imply. One interpretation views $R$ as an {\em independent} fairness notion, and the fact that $R$ is a relaxation of $F$ as a coincidence. Under this interpretation, selection rules consistent with $R$ should select allocations acceptable under $R$ (when there are such allocations), but need not give preference among them to allocations that are also acceptable under $F$. Another interpretation is that $R$ is not an independent fairness notion, but rather that a hierarchy of desirable fairness notions exists, with $F$ higher up than $R$. Under this interpretation, selection rules consistent with $R$ should select allocations acceptable under $F$ (when there are such allocations), and only if no such allocations exist, select other allocations acceptable under $R$ (if there are such allocations). 

Under the independent interpretation, we also treat monotonicity properties of $R$ and $F$ independently. For example it may well be that one of them suffers from the incentive paradox but the other does not. This indeed happens if we take $R$ to be APS and $F$ to be CE (recall Proposition~\ref{pro:CEimplies} that implies that APS is a relaxation of CE). CE suffers from the incentive paradox, whereas APS does not.

Under the hierarchical interpretation, we may want to treat monotonicity properties of $R$ in a way influenced by monotonicity properties of $F$. For example, if $F$ suffers from the incentive paradox, we may view this as implying that $R$ too suffers from the incentive paradox (because $R$ is required to give priority to allocations acceptable under $F$). 

Some of the fairness notions not considered in this paper are relaxations of fairness notions that are considered in this paper. For example, for allocation of indivisible items to agents of equal entitlement, there is the well studied chain of relaxations in which EFX (envy free up to any item) is a relaxation of EF, and EF1 (envy free up to some item) is a relaxation of EFX. (See definitions in~\cite{Budish11, CKMPSW19}.) EF1 and EFX have various extensions to the arbitrary entitlement case. Under the hierarchical interpretation, the fact that WEF suffers from the inversion paradox is inherited also to EF1/EFX type relaxations of WEF.

We hope that if a future study of the monotonicity properties of relaxed fairness notions will be conducted, then that study will explicitly discuss the question of whether it is taking the independent interpretation or the hierarchical interpretation.

\subsection*{Acknowledgements}

This research was supported in part by the Israel Science Foundation (grant No. 1122/22). I thank Moshe Babaioff for many extensive discussions on fair allocations.


\bibliographystyle{alpha}


\newcommand{\etalchar}[1]{$^{#1}$}

\begin{appendix}

\section{MWNSW with concave valuation functions}
\label{sec:MWNSWconcave}

Recall Proposition~\ref{pro:MWNSWdivisible} that shows that for a divisible homogeneous good, MWNSW suffers from the inversion paradox. The proof of that proposition used valuations that are not concave. In contrast, when valuations are concave, MWNSW is monotone.

\begin{proposition}
\label{pro:MWNSWconcave}
    For allocation of a divisible homogeneous good, if all valuation functions are concave, then MWNSW is both individually and globally monotone.
\end{proposition}

\begin{proof}
For simplicity, we first assume that the concave valuations are strictly increasing, an assumption that implies that there is a unique allocation that maximizes WNSW. 

Consider two vectors of entitlements $b >_1 b'$, and the associated MWNSW allocations $A$ and $A'$ (specifying the fraction of the goods that each agent receives). Suppose for the sake of contradiction that $A_1 = A'_1 - \Delta_1$ for some $\Delta_1 > 0$. Then there must be some agent, say agent~2, for which $A_2 = A'_2 + \Delta_2$ for some $\Delta_1 > 0$. We consider two cases.

\begin{itemize}
    \item  $\Delta_1 \le \Delta_2$. 
    
    By optimality of $A$ we have that $$(v_1(A_1))^{b_1}\cdot (v_2(A_2))^{b_2} >  (v_1(A'_1))^{b_1}\cdot (v_2(A_2 - \Delta_1))^{b_2}$$
    By concavity of $v_2$ and because $A_2 \ge A'_2 + \Delta_1$, we have that $\frac{v_2(A'_2 + \Delta_1)}{v_2(A'_2)} \ge \frac{v_2(A_2)}{v_2(A_2 - \Delta_1)}$. Hence, 
    $$(v_1(A_1))^{b_1}\cdot (v_2(A'_2 + \Delta_1))^{b_2} >  (v_1(A'_1))^{b_1}\cdot (v_2(A'_2))^{b_2}$$ 
    By optimality of $A'$ we have that 
    $$(v_1(A'_1))^{b'_1}\cdot (v_2(A'_2))^{b'_2} > (v_1(A_1))^{b'_1}\cdot (v_2(A'_2 + \Delta_1))^{b'_2}$$ 
    The combination of the last two inequalities implies that:
    $$\left(\frac{v_1(A_1)}{v_1(A'_1)}\right)^{b_1 - b'_1}\cdot \left(\frac{v_2(A'_2)}{v_2(A'_2 + \Delta_1)}\right)^{b'_2 - b_2} > 1$$
    The last inequality is a contradiction because $v_1$ and $v_2$ are nondecreasing, $A'_1 > A_1$, $b_1 > b'_1$ and $b'_2 \ge b_2$.

    \item  $\Delta_2 \le \Delta_1$. 

    By optimality of $A'$ we have that 
    $$(v_1(A'_1))^{b'_1}\cdot (v_2(A'_2))^{b'_2} > (v_1(A'_1 - \Delta_2))^{b'_1}\cdot (v_2(A_2))^{b'_2}$$ 
    By concavity of $v_1$ and because $A'_1 \ge A_1 + \Delta_2$, we have that $\frac{v_1(A_1 + \Delta_2)}{v_1(A_1)} \ge \frac{v_1(A'_1)}{v_1(A'_1 - \Delta_2)}$. Hence, 
    $$(v_1(A_1+\Delta_2))^{b'_1}\cdot (v_2(A'_2))^{b'_2} > (v_1(A_1))^{b'_1}\cdot (v_2(A_2))^{b'_2}$$ 
    By optimality of $A$ we have that $$(v_1(A_1))^{b_1}\cdot (v_2(A_2))^{b_2} >  (v_1(A_1 + \Delta_2))^{b_1}\cdot (v_2(A'_2))^{b_2}$$

    The combination of the last two inequalities implies that:
    $$\left(\frac{v_1(A_1)}{v_1(A_1+\Delta_2)}\right)^{b_1 - b'_1}\cdot \left(\frac{v_2(A'_2)}{v_2(A_2)}\right)^{b'_2 - b_2} > 1$$
    The last inequality is a contradiction because $v_1$ and $v_2$ are nondecreasing, $A'_2 < A_2$, $b_1 > b'_1$ and $b'_2 \ge b_2$.
\end{itemize}

We now briefly explain what changes in the proof if valuations are not strictly increasing. Being concave, $v_1$ is strictly increasing up to a point $t_1$, after which it remains constant. If $A'_1 \ge t_1$ then agent~1 did not lose value. Hence, the only problem that might arise is if $A'_1 < t_1$. If $A_1 \le t_1$, the proof works without change. If $A_1 > t_1$, then change the allocation $A$ to a new allocation $B$ in which $B_1 = t_1$ (and some other agent gains a $A_1 - t_1$ piece of the good). This allocation $B$ must have the same WNSW as $A$, as the value received by agent~1 does not decrease. Now the original proof with the pair of allocations $B$ and $A'$ works without change. 
\end{proof}

\section{Infeasibility of natural proper shares even when valuation is external}
\label{sec:external}

Let $MMS_n$ denote the MMS share when there are $n$ agents (the entitlement is $\frac{1}{n}$). For arbitrary entitlement, define $MMS^-$ as the following share. Given entitlement $0 < b < 1$, let $k \ge 2$ be the integer satisfying $\frac{1}{k} \le b < \frac{1}{k-1}$. Then $MMS^- = MMS_k$. Observe that $MMS^-$ is never larger than the pessimistic share, and in instances with equal entitlement, it equals the MMS. Given that the MMS is feasible for instances with  equal entitlement and an external valuation function, one might hope that $MMS^-$ is feasible for instances with arbitrary entitlements and an external valuation function. Unfortunately, this is not true, not even for additive valuations.

\begin{proposition}
    The share $MMS^{-}$ defined above is not feasible, not even if all agents have the same additive valuation function.
\end{proposition}

\begin{proof}
We present here only the construction of an allocation instance. The proof that this instance has no allocation acceptable under $MMS^{-}$ is only sketched, and further details are left to the reader.

    Consider an instance with three agents and entitlements $(b_1, b_2, b_3) = (\frac{1}{2}, \frac{1}{3}, \frac{1}{6})$, and with $2 \cdot 3 \cdot 6 = 36$ items, where item names are vectors with three coordinates, where the entries in the coordinates are taken from $\{1,2\}$, $\{1,2,3\}$, $\{1,2,3,4,5,6\}$ respectively. As the entitlements are inverses of integers, for every agent $MMS^-$ equals the MMS. The intended MMS partition of every agent $i$ is along coordinate $i$. We now explain how to select the values of items in a way that respects these MMS partitions, and moreover, does not allow for an allocation that gives each agent at least her MMS value. The value of each item is the sum of three components. 
    

    The first component is identical for all items and is a fixed constant, which we choose to be 1. With such a constant it will hold that for any two sets of items, if one set is larger than the other, then its value is also larger than that of the other. However, the proof that follows does not depend on this property.  Observe that with this component alone, the MMS values of the agents are 18, 12 and 6 respectively.

    The second component is a perturbation that is item dependent. Its purpose is to leave the MMS partitions unchanged, while ensuring that for every set of items, its value is integer only if it is either one of the bundles of one of the MMS partitions, or a disjoint union of such bundles. To achieve this property we pick a small integer $q$ ($q=2$ may suffice, though picking a larger $q$ makes the proof of correctness easier) and a sufficiently large integer $K$, such as $K = q^{36}$. We partition the items into two groups. The first group, referred to as the correcting group,  contains nine carefully chosen items. Specifically, we may choose for this purpose the items 116, 126, 136, 231, 232, 233, 234, 235, 236. The second group, referred to as the main group, contains the remaining 27 items. For every item $j$ in the main group  we choose an integer $p_j$ that is a different power of $q$ in the range $q^0$ to $q^{26}$. The value of item $j$ is perturbed by adding to it $\frac{p_j}{K}$. Thereafter, the perturbations of the correcting items are computed deterministically in a way that ensures that each bundle in each MMS partition keeps its original value (which is an integer). Item 116 and 126 do this for the first and second bundles (respectively) of agent~2, and items 231, 232, 233, 234 and 235 do so for the first five bundles of agent~3. For example, the value of $p_{231}$ is set to $-(p_{111} + p_{121} + p_{131} + p_{211} + p_{221})$, handling the first bundle in the MMS partition of agent~3.  Afterwards, item 136 does this for the first bundle of agent~1. Finally, item 236 does so for the last bundle of agent~3 (and as the sum of all item values is now 36, also the last bundles of agents~1 and~2 are handled by this).

    At this point, the instance still has an acceptable MMS allocations, but 
    all acceptable allocations have a special form. They are based on the MMS partition of agent~3 that contains six bundles, each of value 6. In these acceptable allocations, agent~1 gets three of these bundles, agent~2 gets two of these bundles, and agent~3 gets the remaining bundle. 
    
    The third component in the valuation function makes allocations such as the above not acceptable. This is done by selecting $\epsilon < \frac{1}{6K}$, and picking six items from different bundles of the MMS partition of agent~3, but belonging to the same bundle for agent~2, and to the same bundle for agent~1. Specifically, we may pick items 231, 232, 233, 234, 235 and 236. For five of these items we reduce their value by $\epsilon$, and for the remaining item (say, 236) we increase its value by $5\epsilon$. The MMS value of agent~3 drops by $\epsilon$, and for the remaining agents it remains unchanged. 

    In this final instance, allocations that were not MMS allocations with respect to the first two components are still not MMS allocations (because $\epsilon$ is small), whereas in allocations that  were MMS allocations with respect to the first two components, now either agent~1 or agent~2 do not receive their MMS value.
\end{proof}

\section{Self maximizing shares}
\label{sec:self-maximizing}


A proper share $s$ is {\em self maximizing} if for every two valuations $v_i$ and $v_j$ and entitlement $b$, there always is a bundle $B$ acceptable according to $v_j$ (namely, $v_j(B) \ge s(v_j,b)$) of $v_i$ value not larger than the share value for $v_i$ (namely, $v_i(B) \le s(v_i,b)$). For a self maximizing share, reporting her true valuation function is a dominant strategy for a risk averse agent that fears receiving the worst among her acceptable bundles -- misreporting a different valuation function will not improve the worst case value of an acceptable bundle. As shown in~\cite{BF22} (who introduced the concept of self maximizing shares, in the context of equal entitlements), self maximizing shares have additional desirable properties, such as being monotone and 1-Lipschitz.

For the purpose of the following proposition, we extend the notion of self maximizing also to non-proper shares, such as the WMMS. (The details of the extension are implicit in the proof of the proposition.)

\begin{proposition}
    APS is self maximizing for every class of valuation functions. Prop and WMMS are not self maximizing, not even for additive valuations.
\end{proposition}

\begin{proof}
    The fact that Prop is not self maximizing (not even for additive valuations) was proved in~\cite{BF22}. For the remaining cases, $v_i$ represents the true valuation function of the agent, whereas $v'_i$ represents an arbitrary different valuation.
    

    To see that the APS is self maximizing, observe that for every price vector, the set of affordable bundles does not depend on the valuation, and there is nothing to be gained by selecting the affordable bundle of highest $v'_i$ value instead of the one of highest $v_i$ value.

    To see that the WMMS is not self maximizing, consider an allocation instance with three agents with additive valuations, and three items. In this instance $b_1 = 0.5$, $b_2 = 0.3$, $b_3 = 0.2$, $v_1(e_1) = 5$, $v_1(e_2) = 3$ and $v_1(e_3) = 1$. The WMMS of agent~1 is $\frac{0.5}{0.2} \cdot 1 = 2.5$, and hence getting only item $e_2$ is acceptable, for a value of~3. However, by misreporting  $v'_1(e_3) = 2$, the WMMS of agent~1 becomes $5$ with respect to $v'_1$, and then every acceptable allocation (with respect to $v'_1$) must give agent~1 a bundle containing the item $e_1$, and hence of $v_1$ value at least $5$.
\end{proof}

\end{appendix}

\end{document}